\documentclass[
twocolumn,
]{aastex631}
\usepackage{amsmath}
\usepackage{graphicx}
\usepackage{xcolor}
\usepackage{xparse}
\usepackage{bm}
\usepackage{latexsym}
\usepackage[
capitalize,
noabbrev
]{cleveref}
\usepackage{tikz}
\usepackage[normalem]{ulem}
\usepackage{paralist}
\usepackage{multirow}

\NewDocumentCommand{\years}{o}{\ensuremath{
\IfNoValueF{#1}{#1 \,}
\mathrm{years}
}}

\NewDocumentCommand{\days}{o}{\ensuremath{
\IfNoValueF{#1}{#1 \,}
\mathrm{days}
}}

\NewDocumentCommand{\months}{o}{\ensuremath{
\IfNoValueF{#1}{#1 \,}
\mathrm{months}
}}

\NewDocumentCommand{\s}{o}{\ensuremath{
\IfNoValueF{#1}{#1 \,}
\mathrm{s}
}}

\NewDocumentCommand{\nT}{o}{\ensuremath{
\IfNoValueF{#1}{#1 \,}
\mathrm{nT}
}}

\NewDocumentCommand{\km}{o}{\ensuremath{
\IfNoValueF{#1}{#1 \,}
\mathrm{km}
}}

\NewDocumentCommand{\au}{o}{\ensuremath{
\IfNoValueF{#1}{#1 \,}
\mathrm{AU}
}}

\NewDocumentCommand{\temp}{o}{\ensuremath{
\IfNoValueF{#1}{#1 \times}
\mathrm{10^5 \; K}
}}

\NewDocumentCommand{\cc}{o}{\ensuremath{
\IfNoValueF{#1}{#1 \;}
\mathrm{cm}^{-3}
}}

\NewDocumentCommand{\pct}{o}{\ensuremath{
\IfNoValueF{#1}{#1 \;}
\%
}}

\NewDocumentCommand{\Rs}{o}{\ensuremath{
\IfNoValueF{#1}{#1 \;}
\mathrm{R_S}
}}

\NewDocumentCommand{\kms}{o}{\ensuremath{
\IfNoValueF{#1}{#1 \;}
\mathrm{km \, s^{-1}}
}}

\NewDocumentCommand{\mWcc}{o}{\ensuremath{
\IfNoValueF{#1}{#1 \;}
\mathrm{mW \cc}
}}

\NewDocumentCommand{\eV}{o}{\ensuremath{
\IfNoValueF{#1}{#1 \;}
\mathrm{eV}
}}

\NewDocumentCommand{\MeV}{o}{\ensuremath{
\IfNoValueF{#1}{#1 \;}
\mathrm{MeV}
}}

\NewDocumentCommand{\nucleon}{s o}{\ensuremath{
\IfNoValueF{#2}{#2 \;}
\IfBooleanTF{#1}{\mathrm{nucleon}}{\mathrm{nuc}}
}}

\NewDocumentCommand{\MeVnuc}{s o}{\ensuremath{
\IfNoValueF{#2}{#2 \;}
\MeV \! /\IfBooleanTF{#1}{\nucleon*}{\nucleon}
}}

\NewDocumentCommand{\Element}{m}{\ensuremath{\mathrm{#1}}}
\newcommand{\He}{\Element{He}}
\newcommand{\C}{\Element{C}}
\newcommand{\N}{\Element{N}}
\newcommand{\Ox}{\Element{O}}
\newcommand{\Ne}{\Element{Ne}}
\newcommand{\Mg}{\Element{Mg}}
\newcommand{\Si}{\Element{Si}}
\newcommand{\Su}{\Element{S}}
\newcommand{\Ca}{\Element{Ca}}
\newcommand{\Fe}{\Element{Fe}}

\NewDocumentCommand{\FIP}{o}{\ensuremath{
\ensuremath{\mathrm{FIP}}\IfNoValueF{#1}{=\eV[11]}
}}

\NewDocumentCommand{\AbSEP}{O{X} O{\Ox}}{\ensuremath{#1/#2}}

\NewDocumentCommand{\PLawExp}{s o}{\ensuremath{b
\IfNoValueF{#2}{\IfBooleanTF{#1}{\approx}{=} #2}}}

\NewDocumentCommand{\MpQ}{o}{\ensuremath{
\IfNoValueTF{#1}{\mathrm{M/Q}}{(\mathrm{M/Q})_{#1}}}}

\newcommand{\vsw}{\ensuremath{v_\sw}}

\NewDocumentCommand{\grate}{o o}{\ensuremath{
\gamma\IfNoValueF{#1}{/\Omega_{#1}}
\IfNoValueF{#2}{= 10^{{#2}}}
}}

\NewDocumentCommand{\gmax}{o}{\ensuremath{
\gamma_\mathrm{max}\IfNoValueF{#1}{/\Omega_{#1}}
}}

\NewDocumentCommand{\kvec}{o}{\ensuremath{
\vec{k} \rho\IfNoValueF{#1}{{_{#1}}}
}}

\NewDocumentCommand{\kpar}{o}{\ensuremath{
{k_\parallel} \rho\IfNoValueF{#1}{{_{#1}}}
}}

\NewDocumentCommand{\kper}{o}{\ensuremath{
{k_\perp} \rho\IfNoValueF{#1}{{_{#1}}}
}}

\NewDocumentCommand{\ani}{s o}{\ensuremath{
R\IfNoValueF{#2}{_{#2}}
\IfBooleanT{#1}{\, [\perp\!/\!\parallel]}
}}

\NewDocumentCommand{\Trat}{s m m o}{\ensuremath{
T_{\IfNoValueF{#4}{{#4};}#2}/T_{\IfNoValueF{#4}{{#4};}#3}
 \IfBooleanT{#1}{\, [\#]}
}}

\NewDocumentCommand{\pbeta}{s o}{\ensuremath{
\beta\IfNoValueF{#2}{_{#2}}
 \IfBooleanT{#1}{\, [\#]}
}}

\NewDocumentCommand{\pbetaR}{o}{\ensuremath{
(\pbeta[\parallel
\IfNoValueF{#1}{;#1}], \ani[#1])
}}

\NewDocumentCommand{\dv}{o}{\ensuremath{\Delta v\IfNoValueF{#1}{_{#1}}}}
\NewDocumentCommand{\ca}{o}{\ensuremath{C_{A\IfNoValueF{#1}{;#1}}}}
\NewDocumentCommand{\dvca}{o o}{\ensuremath{\dv[#1]/\ca[#2]}}

\NewDocumentCommand{\nuc}{o}{\ensuremath{\nu_{c\IfNoValueF{#1}{;#1}}}}
\NewDocumentCommand{\Nc}{o}{\ensuremath{N_{c\IfNoValueF{#1}{;#1}}}}
\NewDocumentCommand{\Ac}{o}{\ensuremath{A_{c\IfNoValueF{#1}{;#1}}}}

\NewDocumentCommand{\tauEXP}{o}{\ensuremath{
\tau_{\mathrm{exp}\IfNoValueF{#1}{;#1}
}}}

\NewDocumentCommand{\tauCC}{o}{\ensuremath{
\tau_{\mathrm{C}\IfNoValueF{#1}{;#1}
}}}

\NewDocumentCommand{\SSN}{o}{\ensuremath{\mathrm{SSN}
\IfNoValueF{#1}{#1}}}
\NewDocumentCommand{\NSSN}{o}{\ensuremath{\mathrm{NSSN}
\IfNoValueF{#1}{#1}}}

\newcommand{\sw}{\ensuremath{\mathrm{sw}}}

\NewDocumentCommand{\qpar}{o}{\ensuremath{
q_{\parallel
\IfNoValueF{#1}{;#1}
}}}

\NewDocumentCommand{\edv}{o}{\ensuremath{
\tilde{E}_{\dv[#1]
}}}

\NewDocumentCommand{\ndays}{o}{
\ensuremath{N_\mathrm{days}{\IfNoValueF{#1}{= {#1}}}}
}

\NewDocumentCommand{\se}{o}{\ensuremath{
S{\IfNoValueF{#1}{_{#1}}}
}}

\NewDocumentCommand{\ab}{o}{\ensuremath{
A{\IfNoValueF{#1}{_{#1}}}
}}

\NewDocumentCommand{\xcorr}{o}{\ensuremath{
\rho
\IfNoValueF{#1}{(#1)}
}}

\NewDocumentCommand{\xhel}{o}{\ensuremath{
\sigma_{c
\IfNoValueF{#1}{,#1}
}
}}

\NewDocumentCommand{\SpecInd}{o}{\ensuremath{\gamma
\IfNoValueF{#1}{_{#1}}}}
\NewDocumentCommand{\QT}{o}{\ensuremath{\mathrm{QT}
\IfNoValueF{#1}{= #1}}}

\NewDocumentCommand{\pten}{o m}{\ensuremath{
\IfNoValueF{#1}{#1 \times }10^{#2}
}}

\NewDocumentCommand{\mean}{}{\ensuremath{\mu}}

\NewDocumentCommand{\var}{}{\ensuremath{\sigma^2}}

\newcommand{\citepossessive}[1]{\citeauthor{#1}'s (\citeyear{#1})}

\NewDocumentCommand{\sect}{o m}{Section~\ref{sec:#2}\IfNoValueF{#1}{ #1}}
\NewDocumentCommand{\eq}{o m}{\cref{eq:#2}\IfNoValueF{#1}{ #1}}
\NewDocumentCommand{\tbl}{o m}{\cref{tbl:#2}\IfNoValueF{#1}{ #1}}

\RenewDocumentCommand{\added}{m}{\textbf{#1}}
\RenewDocumentCommand{\replaced}{m m}{\textbf{#2}}
\RenewDocumentCommand{\deleted}{m}{\relax
}

\RenewDocumentCommand{\added}{m}{#1}
\RenewDocumentCommand{\replaced}{m m}{#2}

\usepackage{multirow}
\usepackage{tablefootnote}

\newcommand{\nsigma}{\ensuremath{1.25 \sigma}}

\NewDocumentCommand{\plotQTSelectionFit}{s}{
\IfBooleanTF{#1}{\begin{figure*}}{\begin{figure}}
\includegraphics[width=\linewidth]{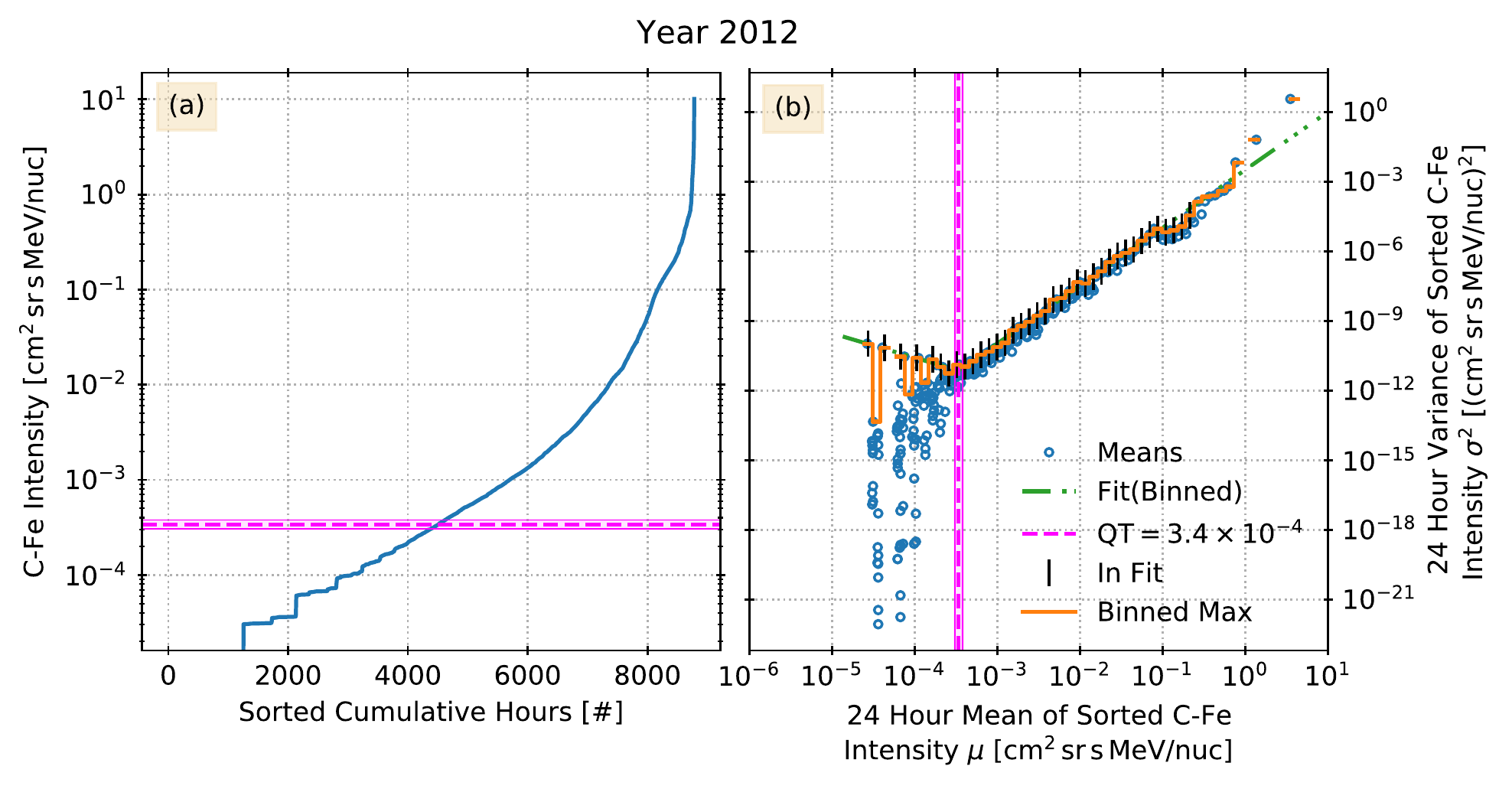}
\includegraphics[width=0.98\linewidth]{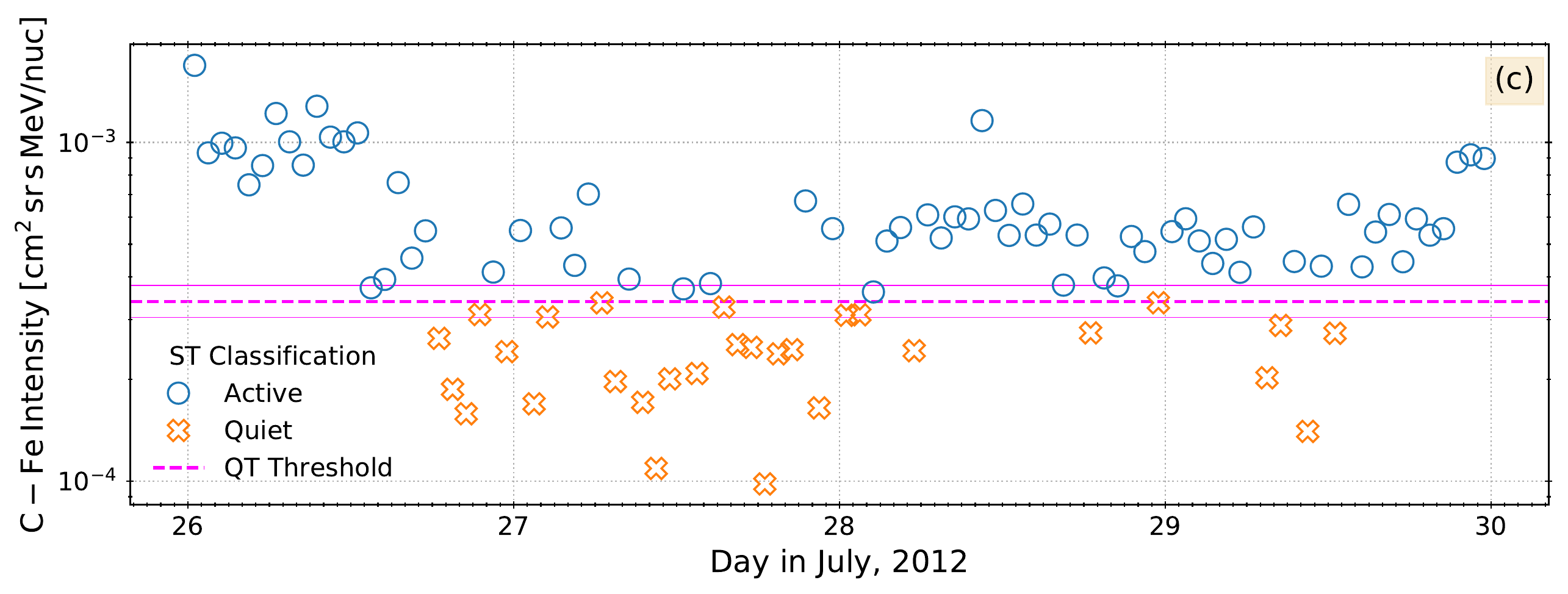}
\caption{
\begin{inparaenum}[\bfseries (a)]
\item The \C\ to \Fe\ intensity over the energy range \MeVnuc*[0.11] to \MeVnuc*[1.29].
\item The variance of the \C--\Fe\ intensity over 24 hour intervals defined in Panel (a) as a function of the corresponding mean, with the maximum of the 24 hour statistics in a fixed number of bins.
\end{inparaenum}
A subset is manually selected for fitting with the maximum of two lines.
Our quiet time threshold (\QT) is the intersection of these two lines and the fits provide a 1$\sigma$ uncertainty on that value (semi-transparent pink).
Data to the left of this threshold corresponds to quiet times.
For reference, panel (a) also includes \replaced{\QT}{the \QT\ threshold} and its $1\sigma$ uncertainty; data below the dashed line corresponds to a quiet time\deleted{ in this panel}.
\textbf{(c)} An example of the cumulative \C\ through \Fe\ intensity time series illustrating the difference between active and quiet times.
Only 7 of 29 quiet time intervals fall within its $1\sigma$ uncertainty.
\label{fig:QT-selection-fit}}
\IfBooleanTF{#1}{\end{figure*}}{\end{figure}}
}

\NewDocumentCommand{\plotQTSelectionSummary}{s}{
\IfBooleanTF{#1}{\begin{figure*}}{\begin{figure}}
\includegraphics[width=\linewidth]{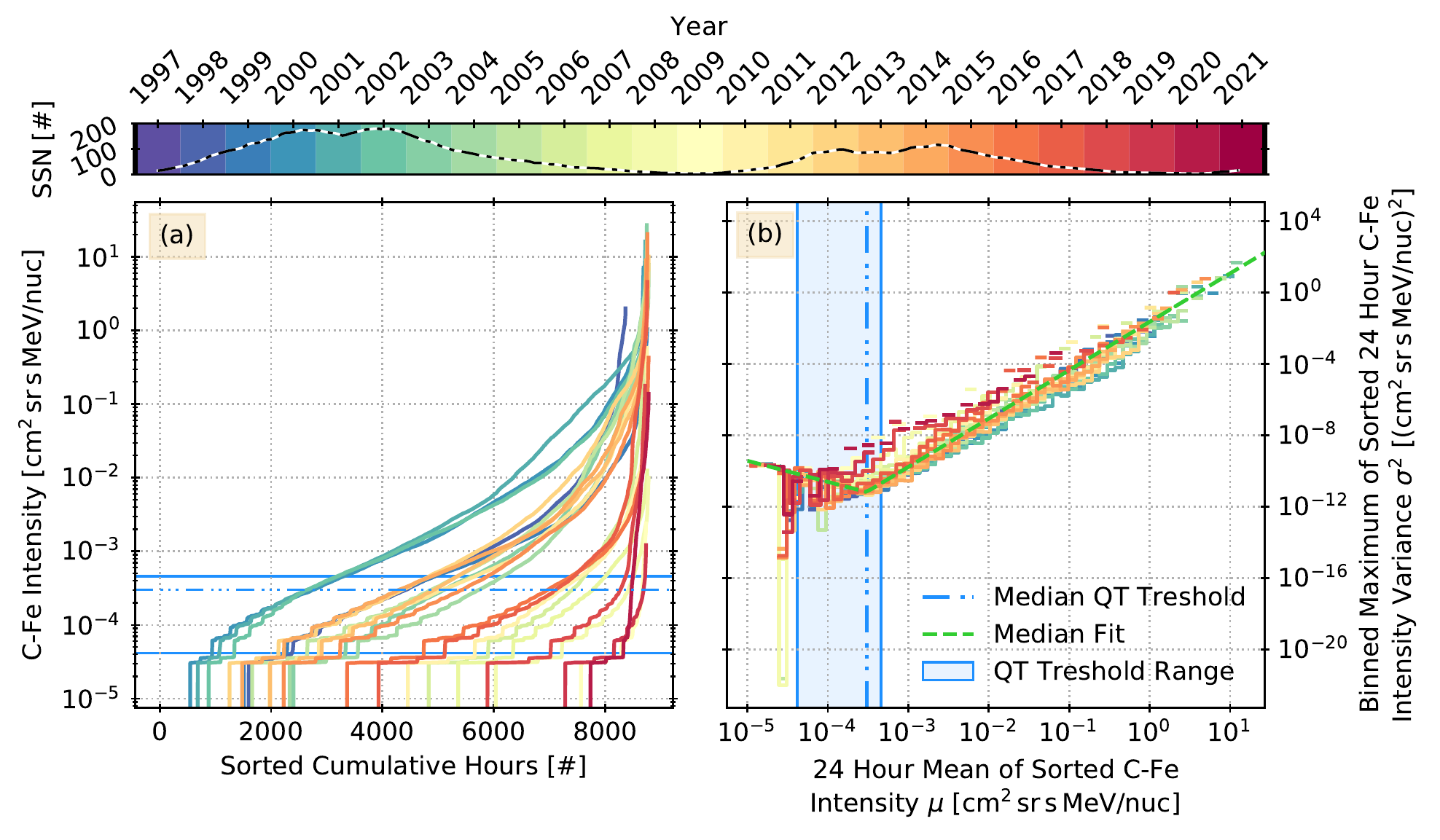}
\caption{\label{fig:QT-selection-summary}
A summary of the \QT\ selection fits per \cref{fig:QT-selection-fit} across all years.
The color bar identifies the year corresponding to each line, \added{with the 13-month smoothed \SSN\ over plotted for context}.
Panels match \cref{fig:QT-selection-fit}.
Panel (b) shows the binned maxima and a summary of the fits across all years.
The \QT\ threshold region on the left panel is completely transparent for visual clarity.
}
\IfBooleanTF{#1}{\end{figure*}}{\end{figure}}
}

\NewDocumentCommand{\plotQTthresholdAndHoursVsTime}{s}{
\IfBooleanTF{#1}{\begin{figure*}}{\begin{figure}}
\includegraphics[width=\linewidth]{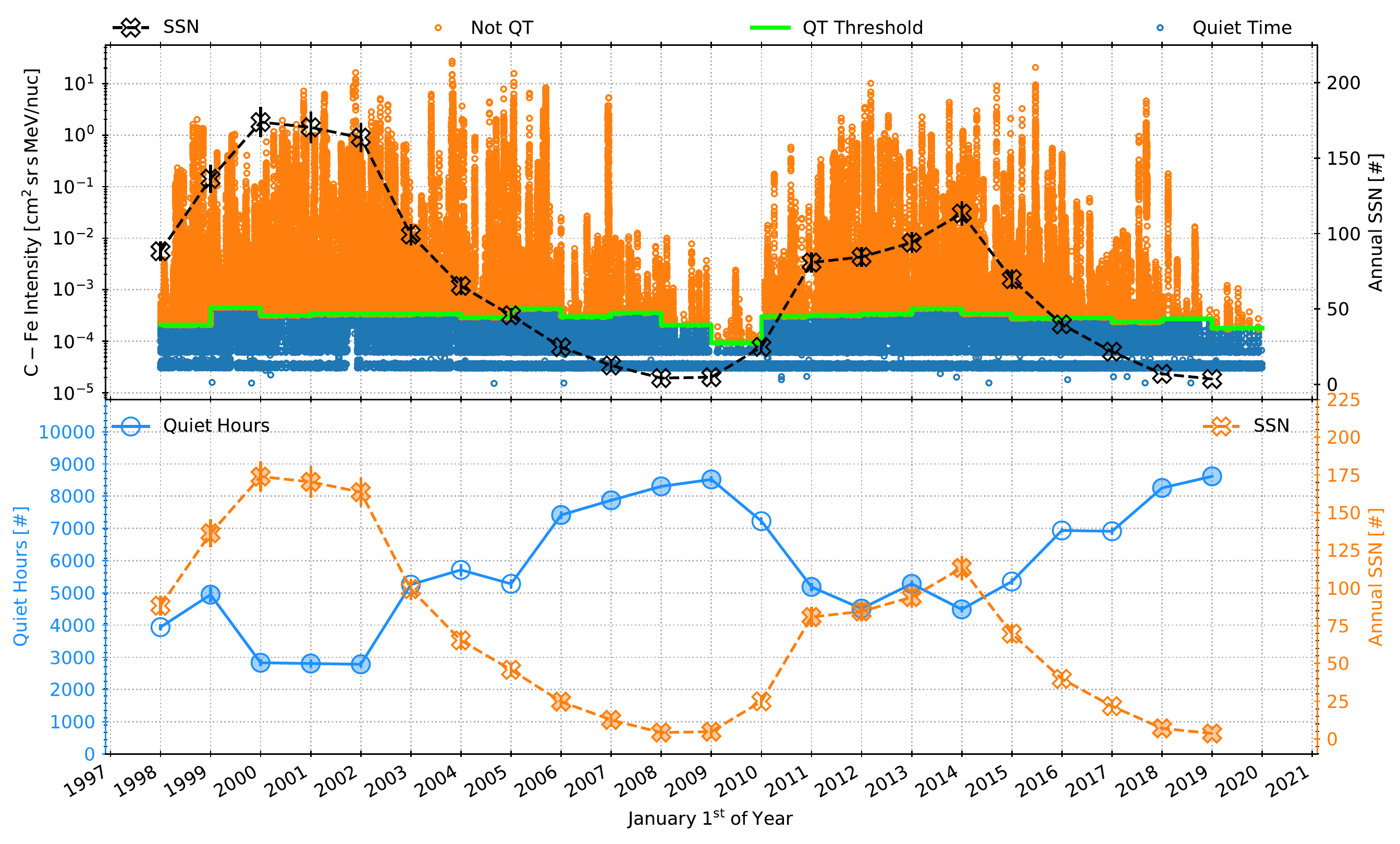}
\caption{\textbf{(a)} Hourly \C--\Fe\ intensity as a function of time.
(\textbf{b}) The left axis plots the annual number of quiet hours.
The right axis \added{in both panels} plots annual \SSN.
\deleted{In both legends, numbers in parentheses indicate the correlation coefficient between the indicated quantity and annual \SSN.}
In the bottom panel, partially filled markers indicate years corresponding to solar cycle extrema, as defined in \cref{sec:annual-ab-extrema}.
\label{fig:QT-threshold-hours}}
\IfBooleanTF{#1}{\end{figure*}}{\end{figure}}
}

\NewDocumentCommand{\plotAnnualAbundance}{s}{
\IfBooleanTF{#1}{\begin{figure*}}{\begin{figure}}
\begin{centering}
\includegraphics[
width=0.95\linewidth
]{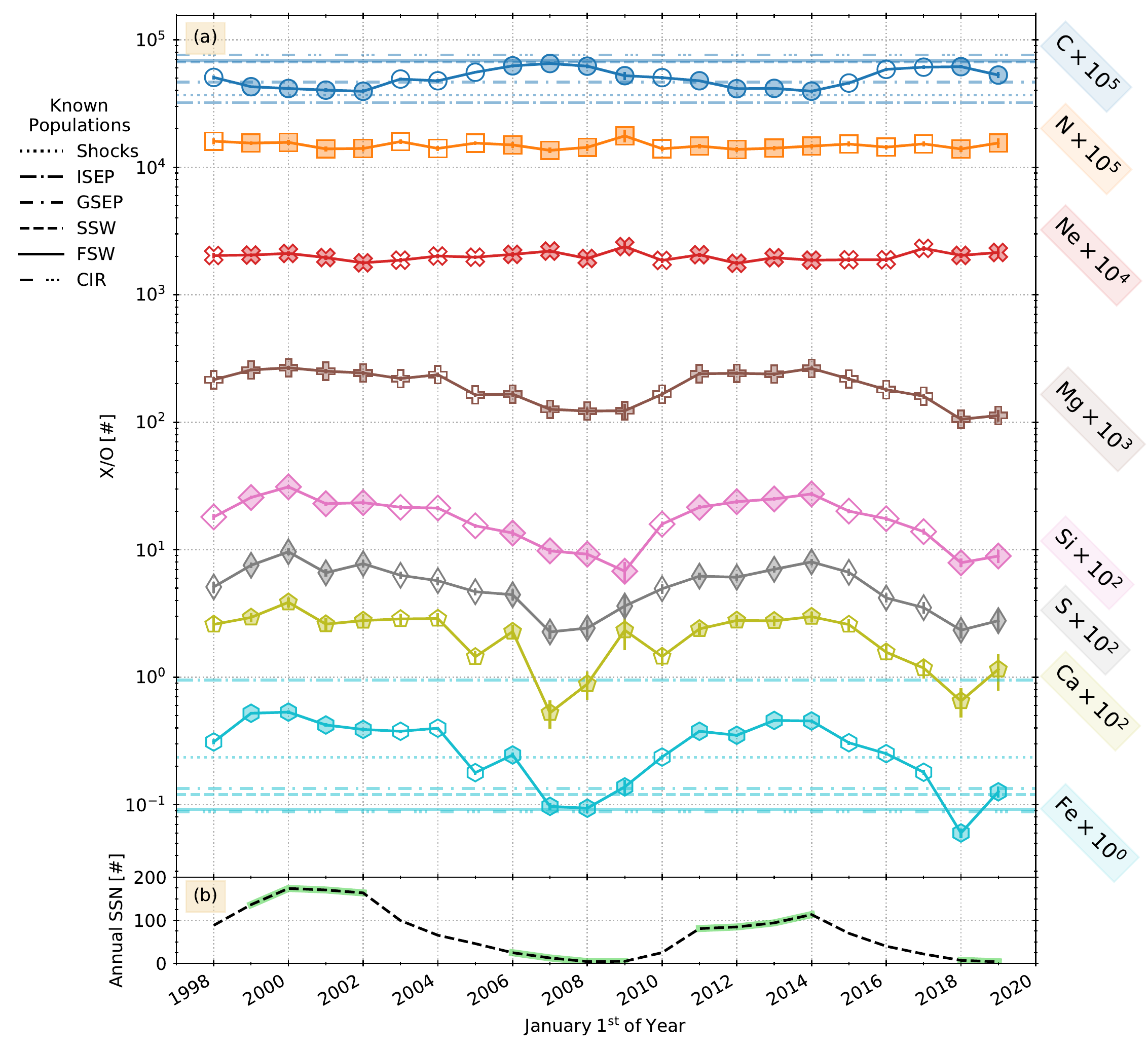}
\caption{\label{fig:annual-ab}
\textbf{(a)} Annual quiet time abundance normalized to oxygen (\AbSEP).
Each species $X$ is labeled on the right hand side of the plot and scaled by the indicated value.
Horizontal lines indicate typical \C\ and \Fe\ abundances for events indicated in the \emph{Known Populations} legend.
\textbf{(b)} Annual SSN, with solar cycle extrema years highlighted in green.
As in \cref{fig:QT-threshold-hours}, \AbSEP\ data occurring during solar cycle extrema are partially filled.
\cref{tbl:Ab-SSN-xcorr} gives the signed correlation of each abundance with SSN \xcorr[\AbSEP,\SSN].
}
\end{centering}
\IfBooleanTF{#1}{\end{figure*}}{\end{figure}}
}

\newcommand{\plotAnnualAbundanceXcorr}{
\begin{figure}
\includegraphics[width=\linewidth]{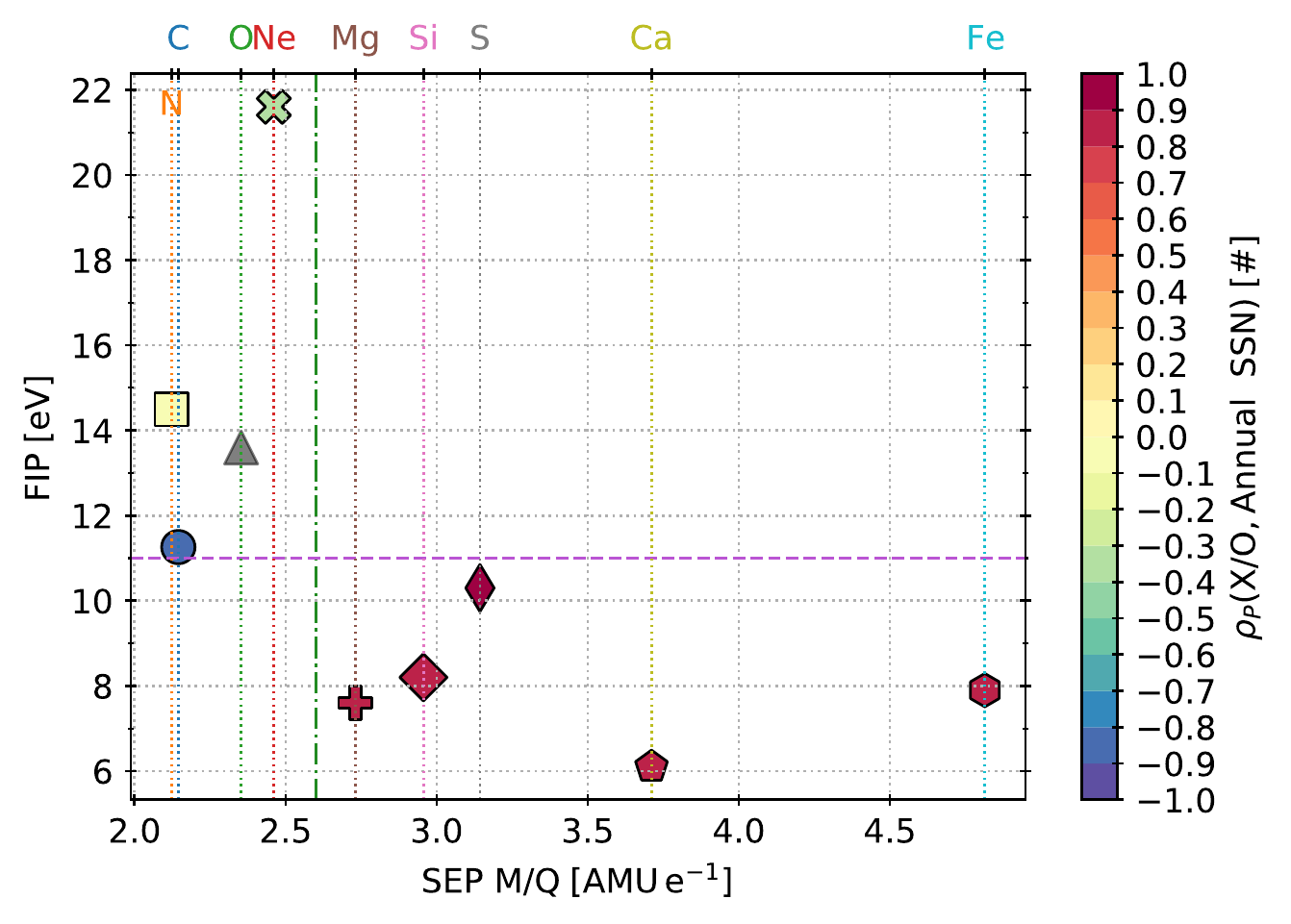}
\caption{\label{fig:annual-ab-xcorr}
The color-coded correlation coefficient $\xcorr[\AbSEP,\SSN]$ between \AbSEP\ and \SSN\ as a function of first ionization potential (\FIP) and SEP \MpQ.
Species are indicated on the top axis and a vertical dotted line connects the species label to its \MpQ.
\Ox\ is indicated in gray for completeness.
The vertical \added{green dash-dotted} line indicates $\MpQ = 2.6$.
The horizontal \added{dashed purple} line indicates \FIP[11].
Considering the impact of acceleration and source region impacts, this figure unexpectedly suggests that \Su\ behaves like a low \FIP\ ion even though $\FIP[\Su] > \eV[10]$.
}
\end{figure}
}

\newcommand{\plotNormAnnualAbundanceMpQ}{
\begin{figure}
\includegraphics[width=\columnwidth]{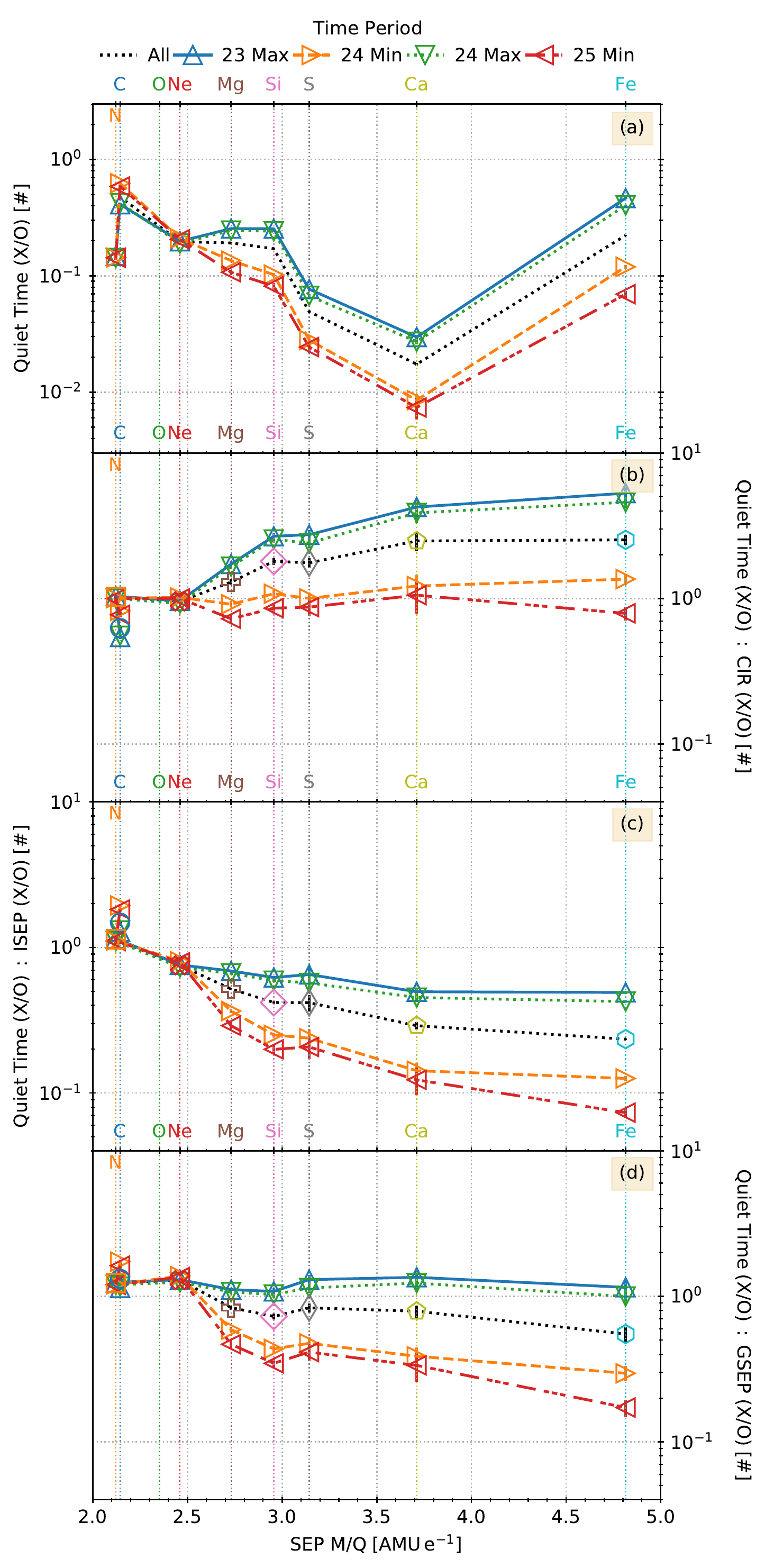}
\caption{\textbf{(a)} Annual quiet time abundance \AbSEP along with these abundances normalized to its known reference abundances in
\textbf{(a)} CIRs,
\textbf{(b)} ISEP, 
\textbf{(c)} GSEP events
as a function of GSEP \MpQ\ for all years along with solar cycle extrema indicated in the legend.
\added{Each species is identified on the top axis.}
\label{fig:norm-annual-ab-MpQ}}
\end{figure}
}

\NewDocumentCommand{\plotScaledMultiSpeciesAbundanceComparisonPlot}{s}{
\IfBooleanTF{#1}{\begin{figure*}}{\begin{figure}}
\begin{centering}
\includegraphics[width=0.808\linewidth]{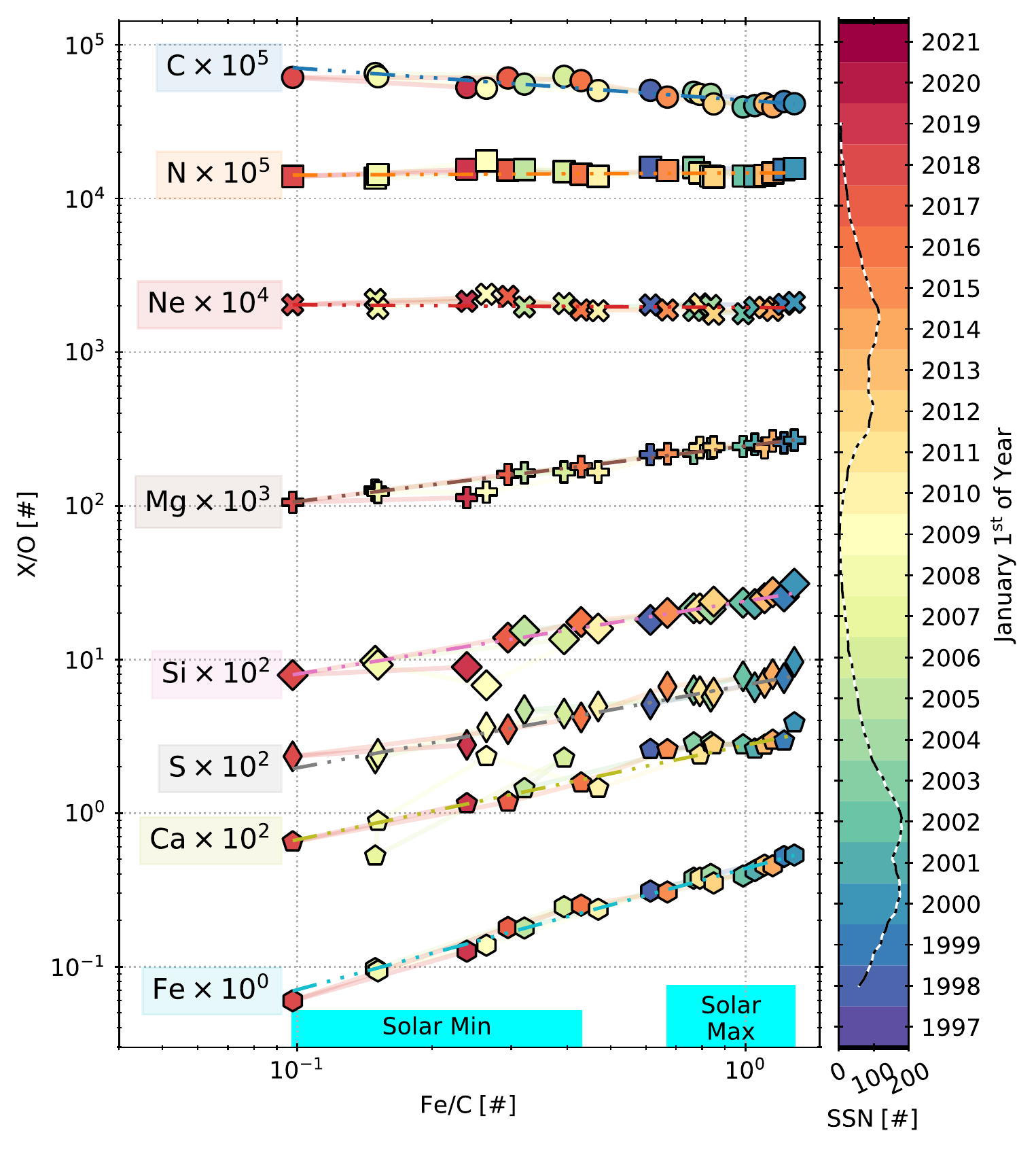}
\caption{The annual abundance \AbSEP\ as a function of \AbSEP[\Fe][\C].
Markers and the connecting segments are colored according to their year; segments are partially transparent.
A power law is fit to each \AbSEP\ as a function of \AbSEP[\Fe][\C] and plotted in the color matching the adjacent label box and \cref{fig:annual-ab}.
\deleted{Changes in \Fe\ likely drive these slopes.}
\label{fig:X/O-vs-Fe/C}}
\end{centering}
\IfBooleanTF{#1}{\end{figure*}}{\end{figure}}
}

\newcommand{\plotAbundanceComparisonPowerLawExponentsByExtrema}{
\begin{figure}
\includegraphics[width=\linewidth]{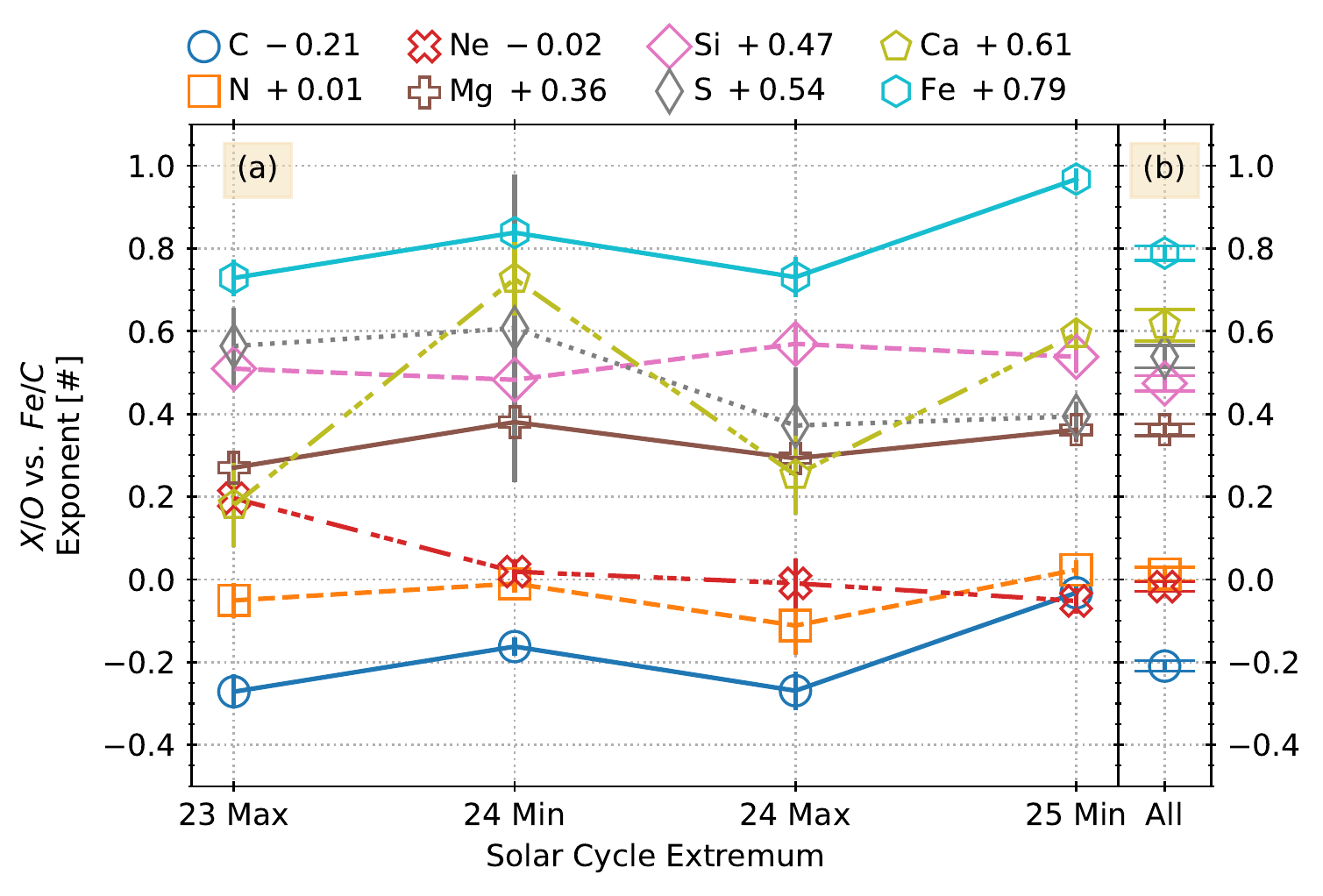}
\caption{\label{fig:X/O-vs-Fe/C-exponents-Extrema}
Power law exponents from fits to $\AbSEP(\AbSEP[\Fe][\C])$.
Panel \textbf{(a)} plots data derived from \cref{fig:X/O-vs-Fe/C} for subsets of the data corresponding to solar cycle extrema.
Panel \textbf{(b)} plots the result for all data from \cref{fig:X/O-vs-Fe/C}.
The legend gives the numerical values from panel (b).
For clarity, only panel (b) shows variability error bars.
While these exponents show some variability, we have too few solar cycle extrema to determine if the variations are sufficiently distinct from the overall average behavior in Panel (b) to be significant.
}
\end{figure}
}

\newcommand{\plotAbundanceComparisonPowerLawExponentsByMpQ}{
\begin{figure}
\includegraphics[width=\linewidth]{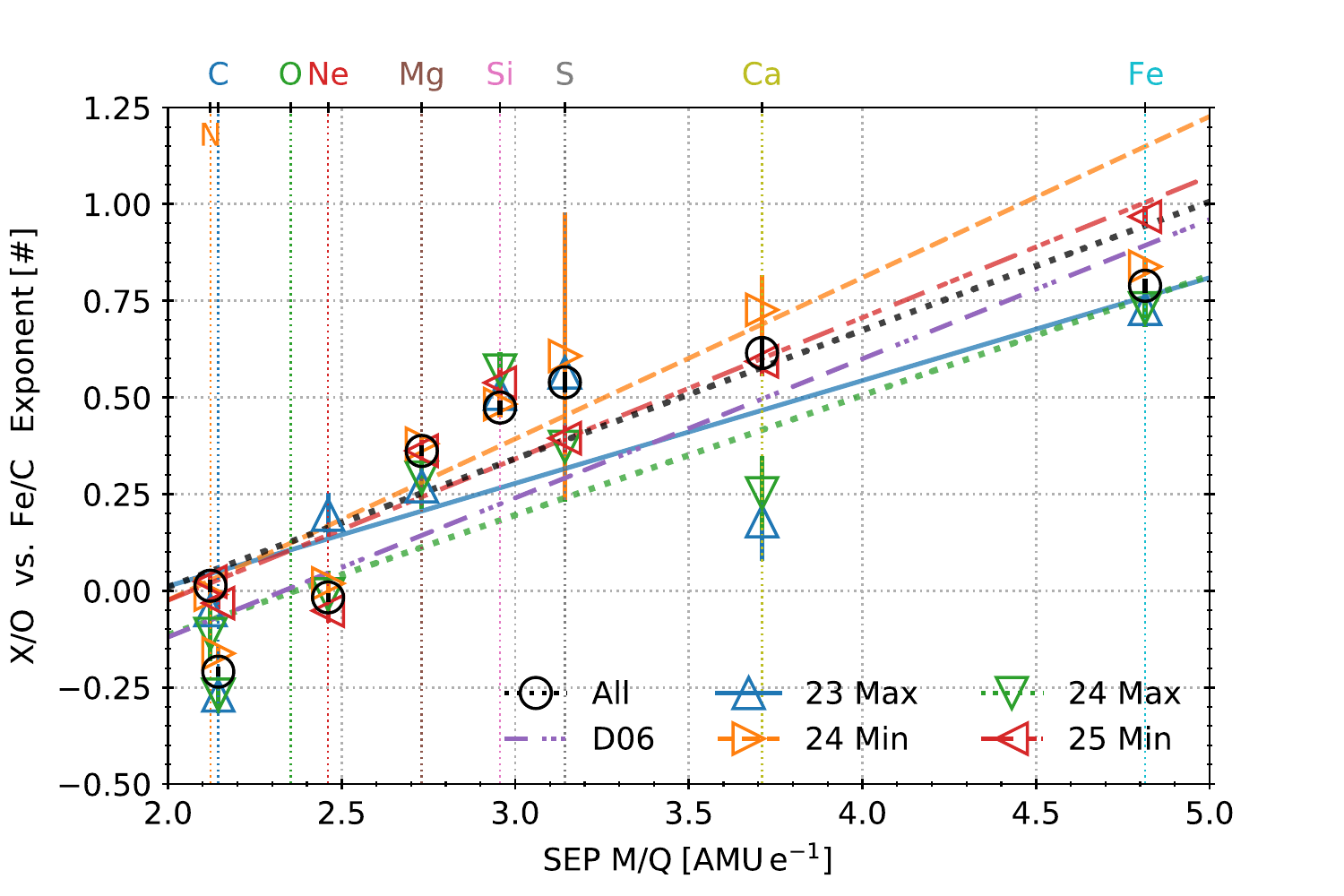}
\caption{\label{fig:X/O-vs-Fe/C-exponents-MpQ}
Power law exponents from fits to \AbSEP\ as a function of \AbSEP[\Fe][\C] as a function of \MpQ\ with lines fit to each data subset indicated in the legend.
The trend line fits are in \cref{tbl:X-vs-Fe}.
The \emph{D06} trend is from the equivalent plot for LSEP events by \citet[Fig.~15]{Desai2006b}.
The consistency across these slopes supports the interpretation that LSEP events accelerate a pre-existing ST pool.
}
\end{figure}
}

\newcommand{\plotFractionationSummaryExtremum}{
\begin{figure}
\includegraphics[width=\linewidth]{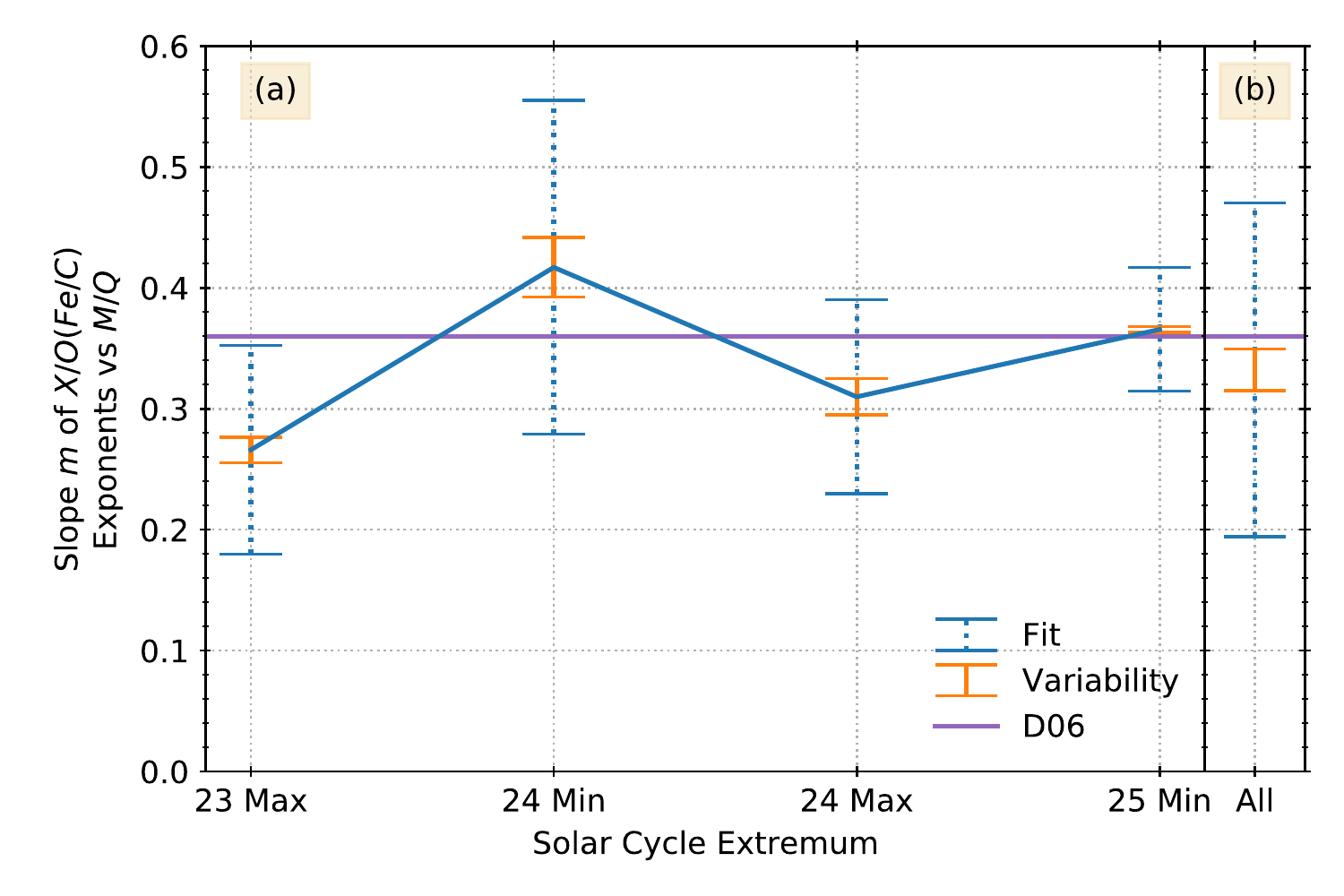}
\caption{\label{fig:Fractionation-Summary}
The fractionation trends derived from \cref{fig:X/O-vs-Fe/C-exponents-MpQ} as a function of \textbf{(a)} solar activity and \textbf{(b)} for all data.
This figure shows that ST \MpQ\ fractionation is independent of solar activity.
}
\end{figure}
}

\newcommand{\eqPLaw}{
\begin{equation}\label{eq:power-law}
\var(\mu) = \var_0 \mu^\epsilon
\end{equation}
}

\newcommand{\eqQTcondition}{
\begin{equation}
\QT(\mu) = \max(\var_1,\var_2) \label{eq:QT-con}
\end{equation}
}

\newcommand{\eqQTthreshold}{
\begin{equation}
\var_1(\QT) = 
\var_2(\QT). \label{eq:QT}
\end{equation}
}

\newcommand{\AnnualAbundanceTable}{
\begin{longrotatetable}
\begin{deluxetable*}{l
rrrrrrrrrrrrrrrr
}
\tablecaption{\label{tbl:annual-ab-data}
Annual abundances normalized to oxygen (\AbSEP) and their uncertainties (Uncert) from \cref{fig:annual-ab}.
All data from ACE/ULEIS over the energy range $\MeVnuc[0.3]$ to $\MeVnuc[1.28]$.
}
\tabletypesize{\scriptsize}
\tablehead{
\colhead{} & \multicolumn{2}{l}{C} & \multicolumn{2}{l}{N} & \multicolumn{2}{l}{Ne} & \multicolumn{2}{l}{Mg} & \multicolumn{2}{l}{Si} & \multicolumn{2}{l}{S} & \multicolumn{2}{l}{Ca} & \multicolumn{2}{l}{Fe} \\
\colhead{Year} & \multicolumn{1}{l}{\AbSEP} & \multicolumn{1}{l}{Uncert} &      \multicolumn{1}{l}{\AbSEP} & \multicolumn{1}{l}{Uncert}&      \multicolumn{1}{l}{\AbSEP} & \multicolumn{1}{l}{Uncert}&      \multicolumn{1}{l}{\AbSEP} & \multicolumn{1}{l}{Uncert}&      \multicolumn{1}{l}{\AbSEP} & \multicolumn{1}{l}{Uncert}&      \multicolumn{1}{l}{\AbSEP} & \multicolumn{1}{l}{Uncert}&      \multicolumn{1}{l}{\AbSEP} & \multicolumn{1}{l}{Uncert}&     \multicolumn{1}{l}{\AbSEP} & \multicolumn{1}{l}{Uncert}
} 
\startdata
1998  &  0.51 &   0.02 &   0.16 &  0.009 &    0.2 &   0.01 &   0.22 &   0.01 &  0.181 &   0.01 &  0.051 &  0.005 &  0.026 &  0.004 &   0.31 &   0.01 \\
1999  &  0.43 &   0.01 &  0.155 &  0.006 &  0.205 &  0.007 &  0.257 &  0.008 &  0.257 &  0.008 &  0.075 &  0.004 &  0.029 &  0.002 &   0.52 &   0.01 \\
2000  &  0.41 &   0.01 &  0.157 &  0.007 &  0.211 &  0.009 &   0.27 &   0.01 &   0.31 &   0.01 &  0.096 &  0.006 &  0.039 &  0.004 &   0.53 &   0.02 \\
2001  &   0.4 &   0.01 &  0.139 &  0.007 &  0.196 &  0.008 &  0.251 &   0.01 &  0.229 &  0.009 &  0.066 &  0.005 &  0.026 &  0.003 &   0.42 &   0.01 \\
2002  &  0.39 &   0.01 &   0.14 &  0.007 &  0.178 &  0.008 &   0.24 &   0.01 &   0.23 &   0.01 &  0.078 &  0.006 &  0.028 &  0.003 &   0.39 &   0.01 \\
2003  &  0.49 &   0.01 &  0.159 &  0.007 &  0.187 &  0.008 &   0.22 &  0.009 &  0.215 &  0.009 &  0.063 &  0.004 &  0.029 &  0.003 &   0.38 &   0.01 \\
2004  &  0.48 &   0.01 &   0.14 &  0.007 &  0.201 &  0.008 &  0.236 &  0.009 &  0.212 &  0.009 &  0.057 &  0.004 &  0.029 &  0.003 &    0.4 &   0.01 \\
2005  &  0.56 &   0.01 &  0.154 &  0.005 &  0.197 &  0.006 &  0.164 &  0.006 &  0.154 &  0.006 &  0.047 &  0.003 &  0.014 &  0.002 &  0.178 &  0.006 \\
2006  &  0.62 &   0.02 &   0.15 &  0.008 &  0.207 &   0.01 &  0.166 &  0.008 &  0.135 &  0.008 &  0.044 &  0.004 &  0.023 &  0.003 &   0.25 &   0.01 \\
2007  &  0.65 &   0.02 &  0.136 &  0.007 &   0.22 &  0.009 &  0.126 &  0.007 &  0.098 &  0.006 &  0.023 &  0.003 &  0.005 &  0.001 &  0.097 &  0.006 \\
2008  &  0.62 &   0.02 &  0.143 &  0.009 &   0.19 &   0.01 &  0.122 &  0.008 &  0.092 &  0.007 &  0.024 &  0.004 &  0.009 &  0.002 &  0.094 &  0.007 \\
2009  &  0.52 &   0.04 &   0.18 &   0.02 &   0.24 &   0.02 &   0.12 &   0.02 &   0.07 &   0.01 &  0.036 &  0.009 &  0.023 &  0.007 &   0.14 &   0.02 \\
2010  &   0.5 &   0.02 &   0.14 &  0.007 &  0.186 &  0.008 &  0.166 &  0.008 &  0.159 &  0.008 &  0.049 &  0.004 &  0.015 &  0.002 &  0.236 &   0.01 \\
2011  &  0.48 &   0.01 &  0.146 &  0.006 &  0.206 &  0.008 &   0.24 &  0.008 &  0.215 &  0.008 &  0.062 &  0.004 &  0.024 &  0.003 &   0.38 &   0.01 \\
2012  &  0.41 &   0.01 &  0.138 &  0.006 &  0.176 &  0.007 &  0.242 &  0.008 &  0.238 &  0.008 &  0.061 &  0.004 &  0.028 &  0.003 &   0.35 &   0.01 \\
2013  &  0.42 &   0.01 &  0.141 &  0.006 &  0.195 &  0.007 &  0.239 &  0.008 &   0.25 &  0.008 &   0.07 &  0.004 &  0.028 &  0.003 &   0.46 &   0.01 \\
2014  &  0.39 &   0.01 &  0.146 &  0.006 &  0.187 &  0.007 &  0.264 &  0.009 &  0.274 &  0.009 &   0.08 &  0.005 &   0.03 &  0.003 &   0.45 &   0.01 \\
2015  &  0.46 &   0.01 &  0.152 &  0.007 &  0.188 &  0.008 &  0.219 &  0.009 &  0.201 &  0.008 &  0.066 &  0.005 &  0.026 &  0.003 &   0.31 &   0.01 \\
2016  &  0.59 &   0.02 &  0.144 &  0.006 &  0.188 &  0.008 &   0.18 &  0.007 &  0.175 &  0.008 &  0.042 &  0.003 &  0.016 &  0.002 &  0.252 &  0.009 \\
2017  &  0.61 &   0.02 &  0.152 &  0.008 &  0.231 &   0.01 &   0.16 &  0.008 &  0.139 &  0.007 &  0.035 &  0.004 &  0.012 &  0.002 &   0.18 &  0.009 \\
2018  &  0.61 &   0.02 &  0.139 &  0.008 &  0.203 &   0.01 &  0.105 &  0.007 &  0.079 &  0.006 &  0.023 &  0.003 &  0.007 &  0.002 &   0.06 &  0.005 \\
2019  &  0.53 &   0.03 &   0.15 &   0.01 &   0.21 &   0.02 &   0.11 &   0.01 &   0.09 &   0.01 &  0.028 &  0.006 &  0.012 &  0.004 &   0.13 &   0.01 \\
\enddata
\end{deluxetable*}
\end{longrotatetable}
}

\newcommand{\AbSSNTable}{
\begin{table}[h]
\centering
\begin{tabular}{c c}
Abundance  & Correlation \\
\AbSEP & Coefficient \\
\hline
\C       & $-0.84$                   \\
\N       & $-0.06$                   \\
\Ne      & $-0.40$                   \\
\Mg      & $+0.90$                   \\
\Si      & $+0.89$                   \\
\Su       & $+0.91$                   \\
\Ca      & $+0.81$                   \\
\Fe      & $+0.89$        \\
\hline          
\end{tabular}
\caption{\label{tbl:Ab-SSN-xcorr}
\replaced{Cross correlation}{Correlation} coefficients for indicated abundances (\AbSEP) with annual \SSN\ \xcorr[\AbSEP,\SSN] from \cref{fig:annual-ab}.}
\end{table}
}

\newcommand{\XvsFeExponentTable}{
\begin{table*}
\begin{tabular}{c ccccccccc}
\hline
{} & \multicolumn{3}{c}{Slope \deleted{$m$}} & \multicolumn{3}{c}{Intercept \deleted{$b$}} & \multicolumn{3}{c}{Cross Correlation} \\
{} &  Value & Uncertainty &   \QT\ &  Value & Uncertainty &   \QT\  &   $\rho$ &      $p$-value &    \QT\  \\
\hline
All      &  0.33 &  0.140 &  0.017 & -0.65 &  0.420 &  0.037 &  0.89 &  \pten[3.0]{-3} &  0.002 \\
23-Max   &  0.27 &  0.090 &  0.011 & -0.52 &  0.280 &  0.041 &  0.76 &  \pten[2.8]{-2} &  0.025 \\
24-Min   &  0.42 &  0.140 &  0.025 & -0.86 &  0.400 &  0.066 &  0.90 &  \pten[2.3]{-3} &  0.005 \\
24-Max   &  0.31 &  0.080 &  0.015 & -0.73 &  0.270 &  0.041 &  0.82 &  \pten[1.2]{-2} &  0.031 \\
25-Min   &  0.37 &  0.050 &     0.002 & -0.76 &  0.170 &  0.010 &  0.94 &  \pten[5.0]{-4} &  0.007 \\
LESP (D06)      &  0.36 &  0.002 &   ... & -0.84 &  0.007 &   ... &  0.97 &  \pten[3.4]{-6} &    ... \\
\hline
\end{tabular}
\caption{\label{tbl:X-vs-Fe}
Parameters from the linear fits in \cref{fig:X/O-vs-Fe/C-exponents-MpQ} along with the correlation coefficient with SEP \MpQ\ and the associated $p$-value.
\QT\ columns give the variability over the \QT\ thresholds, which are smaller than the statistical uncertainty for the \replaced{$\QT = 0$ case}{\QT\ threshold derived in \sect{QT-def}}.
}
\end{table*}
}

\newcommand{\SolarExtremaAbundances}{\footnote{\MeVnuc[0.3 - 1.28]. Averages taken over solar cycle extrema years in this work.}}

\newcommand{\KnownAbundanceTable}{
\begin{table*}[h]
\centering
\caption{\label{tbl:known-abundances}
Representative abundances for known populations along with solar cycle extrema.
Excluding columns (c), (f), and (l), data matches \citet[Table 3]{Desai2006b}.
Although this paper does not analyze He, it is included in this table for completeness.
\added{Where appropriate, energies have been given in the footnotes.}
}
\begin{tabular}{cccccc}
\hline
\hline
{}  &  LSEP\footnote{\MeVnuc[0.1 - 10] with an average value of \MeVnuc[0.38] \citet{Desai2006b}}  &  ISEP\footnote{Impulsive SEP from \MeVnuc[0.32 - 0.45] with or \MeVnuc[\sim\nobreakspace0.385] \citep{Mason2004}}  &  CIR\footnote{Coronating Interaction Region from \MeVnuc[0.32 - 0.45] or \MeVnuc[\sim 0.385] \replaced{\citep{Mason1997,LEICA}}{\citep{Mason2008a,ULEIS}}}  &  ST in Solar Minima\SolarExtremaAbundances &  ST in Solar Maxima\SolarExtremaAbundances \\
        {}  & (a) &  (b)  & (c) & (d) & (e) \\
\hline
\hline
 He  &  $75.0 \pm 23.6$  &  $54 \pm 14$  &  $273 \pm 72$  &  $\ldots$ &  $\ldots$ \\ 
        C  &  $0.361 \pm 0.012$  &  $0.322 \pm 0.003$  &  $0.760 \pm 0.023$  &  $0.6134 \pm 0.0090$ &  $0.4173 \pm 0.0042$ \\ 
        N  &  $0.119 \pm 0.003$  &  $0.129 \pm 0.002$  &  $0.143 \pm 0.005$  &  $0.1434 \pm 0.0037$ &  $0.1453 \pm 0.0022$ \\ 
        O  &  $\equiv 1 \pm 0.02$  &  $\equiv 1 \pm 0.006$  &  $ 1 \pm 0.020 $  &  $ \equiv 1$ &  $ \equiv 1$ \\ 
        Ne  &  $0.152 \pm 0.005$  &  $0.261 \pm 0.003$  &  $0.206 \pm 0.009$  &  $0.2087 \pm 0.0046$ &  $0.1938 \pm 0.0027$ \\ 
        Mg  &  $0.229 \pm 0.007$  &  $0.37 \pm 0.003$  &  $0.148 \pm 0.006$  &  $0.1256 \pm 0.0035$ &  $0.2497 \pm 0.0031$ \\ 
        Si  &  $0.235 \pm 0.011$  &  $0.409 \pm 0.004$  &  $0.095 \pm 0.005$  &  $0.0953 \pm 0.0031$ &  $0.2475 \pm 0.0032$ \\ 
        S  &  $0.059 \pm 0.004$  &  $0.118 \pm 0.015$  &  $0.028 \pm 0.002$  &  $0.0269 \pm 0.0016$ &  $0.0715 \pm 0.0016$ \\ 
        Ca  &  $0.022 \pm 0.002$  &  $0.06 \pm 0.003$  &  $0.007 \pm 0.001$  &  $0.00816 \pm 0.00087$ &  $0.02830 \pm 0.00098$ \\ 
        Fe  &  $0.404 \pm 0.047$  &  $0.95 \pm 0.005$  &  $0.088 \pm 0.007$  &  $0.0984 \pm 0.0032$ &  $0.4298 \pm 0.0044$ \\ 
\hline
\hline
{}  &  SSW \footnote{Slow Solar Wind; \citet{VonSteiger2000b}, \AbSEP[\Ca] from \citet{Wurz2003}} &  FSW\footnote{Fast Solar Wind; \citet{VonSteiger2000b},\AbSEP[\Ca] from \citet{Wurz2003}}  &  Shocks\footnote{\MeVnuc[0.1 - 10] with an average of \MeVnuc[\sim0.75] \citep{Desai2003}}  &  Photosphere\footnote{\citet{Lodders2003}}  &  Corona\footnote{\citet{Feldman2003}}  \\ 
        {}  & (f) & (g) & (h) & (i) & (j) \\ 
\hline
\hline
        He  &  $95.9 \pm 28.8$  &  $72.7 \pm 21.8$  &  $44.4 \pm 14.4$  &  $162 \pm 14$  &  $126 \pm 11$  \\ 
        C  &  $0.67 \pm 0.067$  &  $0.683 \pm 0.068$  &  $0.368 \pm 0.004$  &  $0.501 \pm 0.058$  &  $0.49 \pm 0.056$  \\ 
        N  &  $0.069 \pm 0.021$  &  $0.111 \pm 0.033$  &  $0.142 \pm 0.002$  &  $0.138 \pm 0.022$  &  $0.123 \pm 0.02$  \\ 
        O  &  $ \equiv 1$  &  $ \equiv 1$  &  $ \equiv 1$  &  $\equiv 1 \pm 0.161$  &  $\equiv 1 \pm 0.161$  \\ 
        Ne  &  $0.091 \pm 0.027$  &  $0.082 \pm 0.025$  &  $0.172 \pm 0.003$  &  $0.151 \pm 0.021$  &  $0.191 \pm 0.026$  \\ 
        Mg  &  $0.147 \pm 0.03$  &  $0.105 \pm 0.021$  &  $0.243 \pm 0.004$  &  $0.072 \pm 0.009$  &  $0.224 \pm 0.026$  \\ 
        Si  &  $0.167 \pm 0.034$  &  $0.115 \pm 0.023$  &  $0.213 \pm 0.003$  &  $0.071 \pm 0.007$  &  $0.214 \pm 0.022$  \\ 
        S  &  $0.049 \pm 0.01$  &  $0.056 \pm 0.011$  &  $0.05 \pm 0.001$  &  $0.032 \pm 0.008$  &  $0.032 \pm 0.008$  \\ 
        Ca  &  $0.017 \pm 0.003$  &  $0.0053 \pm 0.0014$  &  $0.022 \pm 0.002$  &  $0.005 \pm 0.0001$  &  $0.013 \pm 0.0002$  \\ 
        Fe  &  $0.12 \pm 0.024$  &  $0.092 \pm 0.018$  &  $0.236 \pm 0.01$  &  $0.061 \pm 0.006$  &  $0.186 \pm 0.017$ \\
\hline
\hline
\end{tabular}
\end{table*}
}

\graphicspath{{./}{figures/}}
\DeclareGraphicsExtensions{.pdf,.png,.jpg,.eps,}

\received{\today}
\submitjournal{ApJ}

\begin{document}

\title{Solar Cycle Variation of \replaced{Suprathermal}{0.3--1.29 MeV/nucleon} Heavy Ion Composition \deleted{and Spectra} during Quiet Times near 1 AU in Solar Cycles 23 and 24}

\shorttitle{Quiet Time Suprathermals in Solar Cycle 23 \& 24}
\shortauthors{Alterman et al.}

\newcommand{\SwRI}{
\affiliation{Space Science and Engineering \\
Southwest Research Institute \\
6220 Culebra Road \\
San Antonio, TX 78238, USA}
}

\newcommand{\APL}{
\affiliation{Applied Physics Laboratory \\
The Johns Hopkins University \\
Laurel, MD 20723, USA}
}

\newcommand{\UTSA}{
\affiliation{Department of Physics and Astronomy\\
 University of Texas at San Antonio\\
 San Antonio, TX 78249, USA}
}

\correspondingauthor{B.\ L.\ Alterman}
\email{blalterman@swri.org}

\author[0000-0001-6673-3432]{B.\ L.\ Alterman}
\SwRI

\author[0000-0002-7318-6008]{Mihir I.\ Desai}
\SwRI
\UTSA

\author[0000-0001-9323-1200]{Maher A.\ Dayeh}
\SwRI
\UTSA

\author[0000-0003-2169-9618]{Glenn M.\ Mason}
\APL

\author[0000-0003-1093-2066]{George Ho}
\APL




\begin{abstract}

We report on the annual variation of quiet-time suprathermal ion composition for C through Fe using Advanced Composition Explorer (ACE)/Ultra-Low Energy Isotope Spectrometer (ULEIS) data over the energy range 0.3 MeV/nuc to 1.28 MeV/nuc from 1998 through 2019, covering solar cycle 23’s rising phase through Solar Cycle 24’s declining phase. 
Our \deleted{novel} findings are
\begin{inparaenum}[(1)]
\item quiet time suprathermal abundances resemble CIR-associated particles during solar minima;
\deleted{and a 70\% gradual solar energetic particle (GSEP) + 30\% impulsive solar energetic particle (ISEP) abundance during solar maxima;}
\item \added{quiet time} suprathermals are \MpQ\ fractionated in a manner that is consistent with \MpQ\ fractionation in large \added{gradual} solar energetic particle events \replaced{(LSEP)}{(GSEP}) during solar maxima; and
\item variability within the \added{quiet time} suprathermal pool increases as a function of \MpQ\ and is consistent with the analogous variability in \replaced{LSEP}{GSEP} events.
\end{inparaenum}
From these observations, we infer that \replaced{there is a feedback mechanism between LSEP events and the suprathermal pool.}{quiet time suprathermal ions are remnants of CIRs in solar minima and GSEP events in solar maxima.}
Coincident with these results, we also unexpectedly show that \Su\ behaves like a low \FIP\ ion in the suprathermal regime and therefore drawn from low \FIP\ solar sources.

\end{abstract}


\section{Introduction \label{sec:intro}} 
\plotQTSelectionFit*
Observations of rare $^3\He$ and $\He^+$ abundances during impulsive solar energetic particle events \citep[ISEP]{Hovestadt1984,Hovestadt1984a,Desai2006}, \added{large} gradual solar energetic particle events \citep[GSEP]{Desai2001,Mason1999a}, and co-rotating interaction regions \citep[CIR]{Chotoo2000} have recorded abundances that are several magnitudes in excess of those observed in the solar wind.
Similarly, observations of $\He^+$ at speeds $2 \vsw$ have been interpreted as interstellar neutrals that have been ionized in the inner heliosphere and picked-up by the solar wind \citep{Mobius1995,Feldman1974b,Blum1970,Holzer1971,Mason2012c}.
Hence they are called ``pickup ions''.
From these observations, the existence of a seed population with energies between the solar wind and energetic particles \citep[few keV to MeV range;][and references therein]{Mason2012c} has been inferred.
Since this population has speeds above the bulk solar wind distribution and below energetic particles, it is referred to
as suprathermal (ST).
\replaced{Follow-up}{Additional} studies of heavy ions further substantiate these observations \citep{Desai2003,Desai2004,Desai2006,Desai2006a,Desai2006b,Desai2007,Filwett2017,Filwett2019}.
As such, ST ions are a likely seed population accelerated in SEP events.
Nevertheless, the mechanism(s) by which the suprathermal energy range is populated remains an open question.

Hypotheses for the \added{source(s) and} \replaced{mechanism}{mechanism(s)} by which the suprathermal energy is populated suggest that either ions are accelerated into this energy range by continuous mechanisms including
\begin{inparaenum}[(1)]
\item turbulence\footnote{\citet{Fisk2006,Gloeckler2008,Fisk2008,Fisk2012a,Fisk2014}}, 
\item velocity fluctuations\footnote{\citet{Fahr2012}}, and 
\item magnetic reconnection\footnote{\citep{Drake2013,Zank2014}}
\end{inparaenum}
or they are remnants of higher energy, discrete processes like 
\begin{inparaenum}[(1)]
\item coronal mass ejection (CME) driven shocks (GSEP events\footnote{\citet{Desai2001,Desai2003,Lario2019,Kahler2019,Mewaldt2012a,Jones1991,Zank2006,Reames1999d,Desai2006a,Dayeh2017,Dayeh2009} \added{Gradual SEP events originally derive their name from the gradual profile of the associated X-ray events. More recently, the name has become associated with the time profile of the energetic particles associated with the event \citep{Reames1999d}.}}), 
\item impulsive flare events (ISEP events\footnote{\citet{Mason2016,Mason2002}}), 
\item and CIRs\footnote{\citet{Allen2019,Mason2008a,Ebert2012a,Fisk1980,Richardson2004,Desai2006a,Dayeh2017,Dayeh2009,Zeldovich2011,Zeldovich2018,Zeldovich2021}}. 
\end{inparaenum}
\citet[Table 2]{LR-Desai} summarize known sources and acceleration mechanisms.

\replaced{Suprathermal ions are difficult for both solar wind and energetic particle instruments to measure because these instruments are either optimized for higher densities at lower energies (solar wind instruments) or lower densities at higher energies (energetic particle instruments) than characteristic of the ST regime.
One path for understanding suprathermals is studying quiet time periods measured by energetic particle detectors. 
In these instruments, quiet times are often selected as the lowest count levels.}{Observations of quiet times can be identified as the lowest count levels in energetic particle detectors}
\replaced{Such methods have lead to a number of key insights about \replaced{the ST pool}{ST ions}.
For example,}{With such methods} \citet{Dayeh2017} showed that the number of quiet hours observed by ACE/ULEIS
\citep{ULEIS} strongly anti-correlates with sunspot number (\SSN).
Multiple studies have also shown that ST abundances (normalized to oxygen, \AbSEP) resemble solar energetic particle (SEP) events-both impulsive and gradual-in solar maxima and CIRs in solar minima \citep{Desai2006a,Dayeh2009,Dayeh2017,Zeldovich2011,Zeldovich2018,Zeldovich2021}.
That the long term changes in ST ions correlate with \SSN\ likely reflects changes in both the sources from which ST ions are drawn and the acceleration mechanisms associated with these sources \citep{Zeldovich2018}.
Whether or not the accelerating conditions are consistent across solar cycles 23 and 24 is still unresolved \citep{Zeldovich2014,Allen2019}.

We study ACE/ULEIS observations from 1998 to 2019 and focus on quiet times. 
We focus on the energy range \MeVnuc[0.3-1.28], hereafter referred to as suprathermals.
Building on \citepossessive{Dayeh2017} techniques, we derive annual quiet time intensity thresholds (QT) with an uncertainty or sensitivity metric.
With this metric, we characterize the sensitivity of our results to our quiet time selection threshold and verify that it has not introduced systematic bias.
We then expand the work of \citet{Desai2006a,Dayeh2009,Dayeh2017} with \C, \Ox, and \Fe\ to study the annual behavior of \C, \N, \Ne, \Mg, \Si, \Su, \Ca, and \Fe\ abundances normalized to oxygen (\AbSEP).

\plotQTSelectionSummary*
The reminder of this paper is organized as follows.
\cref{sec:QT-selection} describes our quiet time selection criteria and its uncertainty metric.
\cref{sec:QT-hours} discusses the annual number of quiet hours.
\cref{sec:ab-annual} presents annual abundances as a function of time.
\cref{sec:annual-ab-extrema} compares them during solar cycle extrema.
\cref{sec:annual-ab-MpQ} studies \MpQ-fractionation of typical ST abundances and how this changes with solar activity.
\sect{ab-comp} examines ST variability as a function of \AbSEP[\Fe][\C] along with what \replaced{these reveal}{this reveals} as a function of \MpQ\ and across solar activity.
\sect{disc} discusses our results \replaced{and contextualizes them in our work, which includes evidence that that LSEP events accelerate an ambient suprathermal population and that \Su\ is a low \FIP\ suprathermal ion.}{and}
\sect{conclusion} concludes.


\plotQTthresholdAndHoursVsTime*
\section{Quiet Time Selection and Quiet Hours \label{sec:QT-def}}
\subsection{Quiet Time Selection \label{sec:QT-selection}}

\citet{Desai2006a} define quiet times in ACE/ULEIS and the Wind Suprathermal through Energetic Particle Telescope (STEP) \citep{Rosenvinge1995} as intervals with a count threshold below a certain value, irrespective of solar activity. \citet[Section 3]{Dayeh2009} build on their work and determine a distinct quiet time threshold for each year by examining the cumulative \C\ through \Fe\ intensity.
They found that the lowest 20\% to 60\% of the intensity values corresponded to quiet times; this percentage depends on the year. \citet[Section 2]{Dayeh2017} refined this method, calculating the mean (\mean) and variance (\var) of this cumulative \C-\Fe\ intensity in consecutive 24-hour intervals when the hourly intensity is sorted in increasing order.\footnote{These intervals do not correspond to chronological days, but rather consecutive 24-hour periods when the data is sorted as described.}
By manually inspecting the data, \citet{Dayeh2017} found an inflection point in \var\ plotted as a function of \mean.
They determined that intervals with \mean\ less than this inflection point's correspond to quiet times.
We refer to the inflection point as \replaced{\QT}{the \QT\ threshold}.

\cref{fig:QT-selection-fit} Panels (a) and (b) plot ACE/ULEIS data over the energy range \MeVnuc*[0.11] to \MeVnuc*[1.29] from 2012 in \citepossessive{Dayeh2017} manner.
Panel (a) plots the cumulative \C-\Fe\ hourly intensity during 2012 in ascending order.
Panel (b) bins the data in Panel (a) into 24-hour intervals and calculates each bin's mean (\mean) and variance (\var), plotting $\var(\mean)$.
Panel (b)'s pink vertical line is the inflection point.
Because \citepossessive{Dayeh2017} thresholds were manually set, determining the sensitivity of this analysis' results to the choice of inflection point becomes arbitrary and their selection criteria remains \emph{ad hoc}.

We build on \citepossessive{Dayeh2017} method, identifying this \QT\ threshold using non-linear fitting.
Along with the \QT\ \replaced{value}{threshold}, our method provides an uncertainty metric, which allows us to quantify the sensitivity of our analysis to the selection of \replaced{\QT}{quiet times}.
We start by generating plots of $\var(\mean)$ for each year in the manner of \cref{fig:QT-selection-fit}.
We have binned \var\ in a fixed number of logarithmically-spaced \mean-intervals and calculated the maximum in each bin plotted as a solid orange line.
For 2018, there are 54 bins, 62 for 2019 and 2020, and 72 otherwise.
We then select a subset of these binned valued indicated with vertical black dashes and fit them in log-space with the maximum
\eqQTcondition
of two power laws
\eqPLaw
subject to the condition that the power laws $\var_1$ and $\var_2$ are equal at the quiet time threshold \QT 
\eqQTthreshold
In effect, this method balances the weight given to the data below and above the inflection point so that we can properly identify the change in slope corresponding to \replaced{\QT}{the \QT\ threshold}.
The $1\sigma$ fit uncertainty for \replaced{\QT}{the \QT\ threshold} is then the interval over which we can test the sensitivity of our results.
That being said, because quiet times inherently involve small numbers, we chose to use the interval $\QT \pm \nsigma$ to provide additional confidence that our fitting methods do not inadvertently include active periods that will dominate our statistics.
\cref{fig:QT-selection-fit} plots this $1\sigma$ fit uncertainty on the \QT\ threshold as a semi-transparent, pink band.
Panel (c) plots an example time series of the cumulative \C\ through \Fe\ intensity during July, 2012.
The QT threshold shows that only 7 of 29 quiet time intervals fall within \QT's $1\sigma$ uncertainty.

\cref{fig:QT-selection-summary} summarizes the \QT\ threshold for all years.
Panel (a) plots the data from all years in the same manner as \cref{fig:QT-selection-fit}.
Panel (b) plots the binned values from all years.
These correspond to the orange line if \cref{fig:QT-selection-fit} (b).
In both panels, the color bar identifies the year corresponding to each line.
While later analysis utilizes the annual \SSN, we over plot the higher time resolution 13-month smoothed \SSN\ (dash-dotted line) on the color bar to provide a time reference instead of the annual \SSN.
In Panel (b), the vertical blue dashed-dotted line is the median \QT\ threshold.
A semi-transparent blue band indicates the range of \QT\ thresholds.
The median of all fit parameters indicates the typical fit in dashed green.
All \SSN\ data is provided by the Solar Information Data Center \citep[SIDC]{sidc,Vanlommel2005}.

\plotAnnualAbundance*
\subsection{Quiet Hours \label{sec:QT-hours}}
\cref{fig:QT-threshold-hours} (a) plots hourly \C--\Fe\ intensities over the energy range \MeVnuc[0.3] to \MeVnuc[1.29] as a function of time.
The \QT\ threshold (solid green) divides quiet time intensities (blue) from other intensities (orange).
Panel (b) plots the number of quiet hours (blue circles) as a function of time; error bars are described below.
Both panel's right axis plots the annual average \SSN\ (``X''); error bars are the standard deviations provided by SIDC.
The Pearson correlation coefficient $\xcorr = 0.51$ between \replaced{\QT}{the \QT\ threshold} and \SSN\ indicates that \replaced{\QT}{the \QT\ threshold} is moderately tied to changes in \SSN.
In contrast, the number of quiet hours strongly anti-correlates with annual \SSN\ at the $\xcorr = -0.95$ level.

To quantify the sensitivity of our results to \replaced{\QT}{the \QT\ threshold selected}, we have calculated 21 logarithmically spaced steps centered on \QT\ over the range $\pm \nsigma$, where $\sigma$ is the \replaced{\QT}{\QT\ threshold} fit uncertainty from \cref{fig:QT-selection-fit}.
We use \nsigma\ to provide additional confidence in the applicability of our method and robustness of our results.
In both panels, \replaced{the these error bars}{we plot the standard deviation of the \QT\ threshold across this $\pm \nsigma$ range as error bars}, which are typically smaller than the markers.


\AbSSNTable
\section{Annual Abundances Over Time and During Cycle Extrema \label{sec:ab}} 
\subsection{Annual Abundances Over Time \label{sec:ab-annual}} 
\cref{fig:annual-ab} Panel (a) plots the annual abundance of quiet time \C, \N, \Ne, \Mg, \Si, \Su, \Ca, and \Fe\ with respect to oxygen (\AbSEP) in the energy range \MeVnuc[0.3] to \MeVnuc[1.29] from 1998 through 2019.
Each species is identified by a label on the right side of the plot and shifted vertically by the indicated value (e.g.~\C\ by $10^{5}$).
As with the intensities in \cref{fig:QT-selection-summary}, these abundances account for and include intervals with zero particle counts.
Error bars indicating the propagated uncertainty of each \AbSEP\ are typically smaller than the markers.
Excluding \Fe\ in 2004 and 2010, the variability as a function of $\QT \pm \nsigma$ range is smaller than the propagated uncertainty and therefore not shown.

\replaced{Panel}{\cref{fig:annual-ab}} (b) plots the annual \SSN.
\cref{tbl:Ab-SSN-xcorr} gives the \deleted{cross} correlation coefficient between \AbSEP\ and \SSN, \xcorr[\AbSEP,\SSN].
In general, \C, \N, and \Ne\ abundances anti-correlate with \SSN\ and the others positively correlate with it.
However, only the positive correlations and Carbon's anti-correlation are strong ($|\xcorr| > 0.6$).
\cref{sec:disc:annual-ab-SSN} discusses the relationship between \xcorr\ and ion \MpQ.

Following \citet{Desai2006b}, \cref{tbl:known-abundances} lists a series of known abundances from a variety of sources and cites the sources for each of these population.
It also includes average abundances from solar minima and maxima, calculated with this paper's data.
Horizontal lines \replaced{identified in the \emph{Known Population} legend}{in \cref{fig:annual-ab}} indicate representative values of \AbSEP[\C] and \AbSEP[\Fe] from the following subset \added{of known populations}:
\begin{compactitem}
\item interplanetary shocks (Shocks),
\item slow solar wind (SSW), 
\item fast solar wind (FSW), 
\item gradual solar energetic particle (SEP) events (GSEP),
\item impulsive SEP events (ISEP), and
\item co-rotating interaction regions (CIR).
\end{compactitem}
\cref{tbl:annual-ab-data} contains the \replaced{associated data.}{data in Panel (a)}.

\plotNormAnnualAbundanceMpQ
\subsection{Abundances During Solar Cycle Extrema \MpQ \label{sec:annual-ab-extrema}}
ST ions are likely \replaced{associate}{associated} with CIRs in solar \replaced{minimum}{minima} \citep{Desai2006a,Dayeh2009,Dayeh2017,Zeldovich2011,Zeldovich2018,Zeldovich2021}.
CIRs are formed when \replaced{fast solar wind streams overtake slow streams}{coronating fast solar wind streams from coronal holes overtake slow wind streams}.
In the outer heliosphere \replaced{, they steepen into shocks}{and occasionally at \au[1], the boundaries of CIRs steepen into forward and reverse shocks}.
Solar wind abundances are governed by their associated sources on the Sun's surface.
Solar sources along with the occurrence rate of flares and CMEs \citep{Webb1994} change with the solar activity cycle.\deleted{During solar minima, the polarity of the solar magnetic field that drives them changes.}
\citet{Alterman2021} showed that the solar cycle driving solar wind abundances changes \days[\sim 250] prior to sunspot minima, likely due to changes in the solar magnetic field \added{that impact solar source regions \citep{McIntosh2014,McIntosh2017,McIntosh2015}}.

To compare abundances across solar cycle extrema, we follow \citet{Zhao2013} and select time periods around each extremum based on the normalized \SSN\ (\NSSN).
\NSSN\ is the \SSN\ in each solar cycle normalized to that cycle's maximum \SSN.
This feature scaling accounts for \SSN's variable amplitude and transforms it into an amplitude-independent clock.
\citet{Allen2019} used a similar method to assign CIRs to solar cycle extrema.
The years we consider to have abundances representative of solar minima have an annual \NSSN[\leq 0.15]. Years with abundances representative of solar maxima correspond to \NSSN[\geq 0.7].
This corresponds to approximately 4 years per extrema and, during years representative of solar minima, primarily selects those from the declining phase of solar activity.
\cref{fig:annual-ab} highlights solar cycle extrema in the bottom panel and partially shades the corresponding \AbSEP\ data points.
Later sections use these intervals to compare ST properties during solar cycle extrema.
By biasing our solar minima selection to the declining phase of solar activity, we ensure that our Minima 24 and 25 intervals are weighted toward ST ions that are generated by a single solar cycle, not an admixture of two \citep{Alterman2021,McIntosh2014,McIntosh2017,McIntosh2015}. \deleted{with their different magnetic polarities and associated features \citep{Alterman2021}.}

\subsection{Abundances as a Function of \MpQ \label{sec:annual-ab-MpQ}}
\replaced{\cref{fig:annual-ab-MpQ}}{\cref{fig:norm-annual-ab-MpQ}} (a) plots \AbSEP\ during solar cycle extrema as a function of GSEP \MpQ.
We use the same average GSEP charge states as \replaced{\citet[Section 5.2]{Desai2006b}.}{Section 5.2 of \citet{Desai2006b}.
Each element's symbol is indicated on the top axis.}
As expected from \cref{fig:annual-ab}, variability of any single data point with the threshold selected in Section 2.1 is smaller than the statistical uncertainty associated with that data point.
The plot \deleted{also}shows a clear trend with \added{increasing} \MpQ: 
\replaced{\AbSEP\ decreases from \Ne\ to \Ca\ in solar minima and \Si\ to \Ca\ in solar maxima.}{the abundances decrease from \C\ to \Ca, but \Mg\ and \Si\ lie above this line.
The abundances then increase from \Ca\ to \Fe.}
During both solar cycle extrema, \AbSEP[\Fe] is more similar to \AbSEP[\Mg] and \AbSEP[\Si] than \AbSEP[\Su] or \AbSEP[\Ca].
The trend for solar maxima-representative abundances in Panel (a) is also qualitatively similar to \added{0.14, 1.1, and 10 \MeVnuc\ ions reported by} \citet[Fig.~6]{Mewaldt2003} .

\plotScaledMultiSpeciesAbundanceComparisonPlot*
\added{ST ions likely suffuse the heliosphere \citep{Tsurutani1985b,Desai2003,Wiedenbeck2003}.
They are variously associated with CIRs \citep{Chotoo2000}, ISEPs \citep{Hovestadt1984,Hovestadt1984a,Desai2006}, and GSEPs \citep{Desai2001,Mason1999a}, depending on the phase of solar activity.
In particular, prior observations suggest that CIRs are the dominant source of ST ions during solar minima \citep{Desai2006a,Dayeh2009,Dayeh2017,Zeldovich2011,Zeldovich2018,Zeldovich2021} and both ISEP and GSEP events dominate ST sources during solar maxima \citep{Dayeh2017,Dayeh2009,Desai2006a}.
}
To quantify the similarity of suprathermal abundances to the known populations in \cref{tbl:known-abundances}, we have generated plots of quiet time \AbSEP\ normalized to \replaced{known abundances}{them} $\mathrm{Quiet \, Time} \, \left(\AbSEP\right) : \mathrm{Known} \left(\AbSEP\right)$ as a function of \MpQ, averaging over each solar cycle extremum and all years from \cref{fig:QT-threshold-hours}.
\replaced{\cref{fig:annual-ab-MpQ}}{Panel} (b) plots the CIR case.
With the exception of carbon, all suprathermal abundances resemble their CIR abundance in solar minima.
Given that \C\ is known to be enhanced in CIRs and the solar wind with respect to the photosphere and corona \citep{VonSteiger2000b,Lodders2003,Feldman2003,Mason1997},
this \replaced{suggests that \AbSEP[\C] is unique in comparison with the other low \MpQ\ abundances}{is not unexpected} \citep{Desai2006a,Dayeh2009,Dayeh2017}.
\added{Panels (c) and (d) plot \AbSEP\ normalized to representative ISEP and GSEP abundances, respectively.
During solar maxima, normalizing quiet time \AbSEP\ to the representative ISEP does not remove the  \MpQ\ trend, but normalizing to the GSEP abundances does.}


\section{Abundances as a Function of \AbSEP[\Fe][\C] \label{sec:ab-comp}} 
To characterize the variability of ST abundances, \cref{fig:X/O-vs-Fe/C} plots each species' abundance \AbSEP\ as a function of \AbSEP[\Fe][\C].
We follow \citet{Reames1994,Mason2004} and take, ``\AbSEP[\Fe][\C] as the most sensitive indicator of the general enrichment of heavier elements relative to lighter ones.'' \citep{Reames1994}
\deleted{As with \cref{fig:QT-selection-summary},} Data points and connecting line colors correspond to the year of observation.
Marker shapes for each species match prior figures and they are connected to aid the eye.
Species are indicated as in \cref{fig:annual-ab}.
Propagated error on the abundances is, excluding \Ca, at most the marker size and all are excluded from the plot for visual clarity.
Dash-dotted lines are power law fits to the data.
Blue boxes indicate the range of values corresponding to each solar cycle extrema.
In general, these trends show cyclic behavior with lower \AbSEP[\Fe][\C] and the correlated \AbSEP\ values during solar minima and the higher \AbSEP[\Fe][\C] values during solar maxima.
Unlike $^3\mathrm{He}$ \added{in ISEP events} \citep{Mason2004}, these ST observations are not uniformly distributed with \added{$\ln\left(\AbSEP[\Fe][\C]\right)$}, but rather show a large spread during solar minima and a concentration during solar maxima.
\deleted{That the distribution of abundances with \AbSEP[\Fe][\C] differs from $^3\He$ rich ISEP events further suggests that ISEP events are not the sole source of the suprathermal ions studied in this paper.}
These slopes are consistent with the trends in \cref{fig:annual-ab} \added{and, as can be inferred from \cref{fig:annual-ab}, \AbSEP[\C] is the exception to this trend showing an anti-correlation.}
\AbSEP[\Ca] is the most significant example, showing a weak gradient during solar maximum and periods of high \AbSEP[\Fe][\C].

\plotAbundanceComparisonPowerLawExponentsByExtrema
\cref{fig:X/O-vs-Fe/C-exponents-Extrema} plots the power law exponents derived in the same manner as \cref{fig:X/O-vs-Fe/C} for 
\begin{inparaenum}[(a)]
\item each solar cycle extremum as a function of time along with
\item all the data.
\end{inparaenum}
Individual plots for deriving each slope in panel (a) are not show for \added{lack of} space.
The data are plotted at the \replaced{derived \QT\ threshold}{value corresponding to the \QT\ threshold derived in \sect{QT-def}} and error bars representing the sensitivity to the \QT\ threshold's uncertainty are typically smaller than the markers.
Excluding \Ca\ along with possibly \Su\ and \Ne, the power law exponents \PLawExp\ are roughly consistent across the cases plotted.
\C\ has the smallest \replaced{\PLawExp[-0.21]}{exponent of -0.21} and \Fe\ the largest \replaced{\PLawExp[0.79]}{0.79}\deleted{, which agrees with the range of values observed in \cref{fig:annual-ab}}.
In contrast to the \replaced{rough}{possibly cyclic} variation of that most species' exponents demonstrate with solar activity, \Ne\ and \Su\ both decrease from solar Maximum \replaced{24}{23} to Minimum 25.
However, more than four solar cycle extrema are required to determine if this difference is due to random fluctuations or is indicative of a true difference between these species and the rest of those studied.
The large change in \Ca's slopes are due to a flattening of \AbSEP[\Ca] during solar maxima observed in \cref{fig:X/O-vs-Fe/C}, which drives these slopes down.
While \C, \N, \Mg, \& \Fe\ show some consistent variability from solar minima to maxima, the power law exponents are markedly more consistent and show less overall variation across cycle extrema.

\plotAbundanceComparisonPowerLawExponentsByMpQ
To characterize the variability of \added{quiet time} ST \replaced{populations}{ions studied in this paper} as a function of mass-per-charge, \cref{fig:X/O-vs-Fe/C-exponents-MpQ} plots these exponents \deleted{in \cref{fig:X/O-vs-Fe/C-exponents-Extrema} }as a function of \MpQ.
In general, the slopes increase with \MpQ, implying that larger \MpQ\ \added{quiet time} ST ions have a more variable enrichment pattern across the solar cycle.
During solar maxima, \Ca\ outliers are likely due to the flattening noted in \cref{fig:X/O-vs-Fe/C}.
Broadly, there may also be a change in the variability trend at \Su, where the change in slopes is steeper for $\MpQ < \MpQ(\Su)$ and shallower otherwise.
\replaced{However, more than two data points at $\MpQ > \MpQ(\Su)$ are necessary to make any definitive inferences.
As such,}{To characterize the general nature of this trend,} we have fit the data in \cref{fig:X/O-vs-Fe/C-exponents-MpQ} with a line, \replaced{as it is the simplest representation of this trend.}{reserving more detailed analysis for future study.}
We have also plotted the trend from \citet[Fig.~15]{Desai2006b} for \replaced{LSEP events}{GSEP event at \MeVnuc[0.38]}.
Broadly, the trends during solar maxima are consistent with the \replaced{LSEP}{GSEP} trend.
\deleted{As \citepossessive{Desai2006b} observations \added{of individual GSEP events} cover one solar cycle, our observations suggest that event-to-event variability in \added{their} \replaced{LSEP}{GSEP} events is consistent with the long-term variability of ST abundances \added{in quiet times}.}
\deleted{As such, we take these observations as evidence that LSEP events accelerate ions out of a ST pool.}

\plotFractionationSummaryExtremum
\cref{fig:Fractionation-Summary} plots the fractionation slopes from \cref{fig:X/O-vs-Fe/C-exponents-MpQ} (a) across solar cycle extrema and (b) for all data.
Here, we plot the \replaced{\QT\ value}{value corresponding to the threshold derived in \sect{QT-def}} with its fit uncertainty \added{with dotted blue error bars} and variability with \QT\ variability with solid orange error bars.
\added{The variability error bars are calculated as in \sect{QT-hours}.}
D06 is the slope from \citet[Fig.~15]{Desai2006b}.
Two observations stand out.
\begin{inparaenum}[(1)]
\item The fit uncertainty is markedly larger than the variability of our results due to \QT\ threshold selection.
\item The horizontal line from \citet{Desai2006b} (nearly) intersects the \emph{Fit} error bars for all four solar cycle extrema.
\end{inparaenum}


\section{Discussion \label{sec:disc}}

Quiet times are periods when the intensity of the suprathermal population is low.
While the suprathermal intensity naturally varies with solar activity, quiet times are nevertheless present throughout the solar cycle \citep{Desai2006a,Dayeh2009,Dayeh2017,Zeldovich2011,Zeldovich2018,Zeldovich2021}.
We study these suprathermal ions in quiet times across solar activity to characterize their \replaced{implications for more active intervals, e.g.~when a CME-driven shock accelerates these suprathermals into energetic particles.}{origin.}


\subsection{Quiet Time Selection and Annual Quiet Hours \label{sec:disc:qt-selection}} 
\deleted{Suprathermals occupy an energy range for which \emph{in situ} measurements require careful data selection, which is often tailored to a specific study.}
\added{The definition of quiet times is still unsettled in the literature and remains \emph{ad hoc}.}
For example, \citet{Zeldovich2014,Zeldovich2018,Zeldovich2021} select data based on multiple criteria from several instruments across a few spacecraft.
In contrast, \citet{Desai2006a,Dayeh2009,Dayeh2017} set thresholds for the \Fe\ or total \C\ through \Fe\ intensity and allow for \deleted{suprathermal }intensities that report zero counts, i.e.~below the instrument detection threshold.
Both selection methods rely on one or more manually identified thresholds without a metric to quantify the sensitivity of the results to that selection.
\deleted{In short, the definition of quiet times is still unsettled in the literature \added{and is \emph{ad hoc}}.}
\replaced{Because of the \emph{ad hoc} nature of most quiet time definitions}{As such}, quiet time selection necessarily leads to \replaced{improper}{inconsistent} labeling of some intervals as quiet or not.
For the purposes of this study, a false positive would be a not-quiet interval labeled as quiet and a false negative would be a quiet interval labeled as not-quiet.
Quantifying the sensitivity of our result to our quiet time selection criteria provides evidence that such false positives and false negatives do not impact our results.

\cref{sec:QT-selection} utilizes non-linear fitting to identify the \QT\ threshold \replaced{first}{using the definition} described by \citet{Dayeh2017} and provide an accompanying 1$\sigma$ fit uncertainty.
To study the sensitivity of our results to the fit \QT\ threshold, we have calculated 21 uniformly spaced steps over the range ($\QT \pm \nsigma$), centered on the \QT\ fit result.
We repeat our analysis for each of these 21 thresholds and, where applicable, representing the variation as the standard deviation of our results as error bars centered on the \QT\ \replaced{value}{threshold} \citep{Alterman2019}.

\cref{sec:QT-hours} analyzes each year's \QT\ threshold and the number of quiet hours, both as a function of time.
In general, the \QT\ threshold is weakly correlated in time with \SSN\ ($\xcorr = 0.51$).
However, the number of quiet hours anti-correlates with \SSN\ at the $\xcorr = -0.95$ level, likely because solar activity drives the occurrence of non-quiet time periods.
It is unsurprising that our \deleted{cross-}correlation is stronger than that derived by \citet{Dayeh2009,Dayeh2017} because their time series are 13 \added{years\footnote{Approximately solar cycle 23}} and 17 years\footnote{\hphantom{0ex}\added{Approximately cycle 23 into the declining phase of cycle 24.}}, where as ours covers 23 years and longer time series provide additional data with which to characterize any trend or lack there of.
In both the case of the \QT\ threshold and the annual number of quiet hours, our results were insensitive to changes in the \QT\ threshold across the $\pm \nsigma$ range.


\subsection{Annual Abundances: Suprathermal Sources \label{sec:disc:annual-ab-MpQ}}

Studying 41 CIRs, \citet{Mason2008a} infer that energetic particles (EPs) associated in CIRs are drawn from a ST pool composed of solar wind, pickup ions, and SEP remnants.
\citet{Filwett2017,Filwett2019} came to similar conclusions about ST ions accelerated by CIRs.
\cref{fig:norm-annual-ab-MpQ} (b) shows that normalizing the quiet time \AbSEP\ to an abundance characteristic of CIRs observed at \MeVnuc[0.385] removes the \MpQ-dependence during solar minima.

Studying 72 interplanetary shocks at \MeVnuc[0.75] in solar cycle 23, \citet{Desai2003} suggest that the suprathermal pool from which shocks accelerate energetic particles is composed of 70\% GSEP and 30\% ISEP remnant material, both from the \MeVnuc[5-12] range.
Studying 64 LSEP events from solar cycle 23 at \MeVnuc[0.38], \citet{Desai2006b} suggest that GSEP events are \MpQ-fractionated in because of the ambient \MeVnuc[0.38] suprathermal seed population from which they accelerate ions and this suprathermal population is itself primarily drawn from \MeVnuc[0.385] ISEP events.
\cref{fig:norm-annual-ab-MpQ} (c) and (d) show that normalizing quiet time \AbSEP\ to reference abundances from ISEP does not remove the \MpQ-dependence, but normalizing it to a GSEP-representative abundance does remove the \MpQ-dependence during solar maxima.
A mixture of GSEP and ISEP abundances may not be necessary to characterize the quiet time \AbSEP\ fractionation and \citepossessive{Desai2003} observation that suprathermals are composed of 70\% GSEP + 30\% ISEP remnant particles may be due to their comparison of \MeVnuc[0.75] ions with reference abundances at \MeVnuc[5-12].

Suprathermal ions have been observed to suffuse the heliosphere \citep{Tsurutani1985b,Desai2003,Wiedenbeck2003}.
Long term studies of ST ions across solar cycles 23 and 24 also related differences in ST observations at \au[1] to different solar source regions \citep{Zeldovich2018,Zeldovich2014}.
That quiet time \AbSEP\ is most similar to CIRs during solar minima and GSEP abundances during solar maxima further substantiates observations their dominant source changes with solar activity.
That these observations are taken during quiet times also substantiates that suprathermals suffice the heliosphere across solar activity.
The presence of \MpQ-fractionation points to the impact of acceleration on the observed abundances.
That the fractionation disappears when normalized to CIR abundances in solar minima and GSEP abundances in solar maxima may imply that SEPs simply decrease in intensity as they spread out through the heliosphere.


\subsection{Abundances as a Function of \AbSEP[\Fe][\C]: Variability of the \added{Quiet Time} Suprathermal Pool \label{sec:disc:ab-comp}} 

\cref{fig:X/O-vs-Fe/C} plots each species' abundance as a function of \AbSEP[\Fe][\C].
Unlike analogous observations during $^3\He$-rich ISEP events \citep{Mason2004}, these suprathermal observations are not normally distributed with $\ln\left(\AbSEP[\Fe][\C]\right)$.
This suggests ISEP events are not the sole source of the suprathermal ions studied in this paper.

\cref{fig:X/O-vs-Fe/C-exponents-Extrema} plots exponents from the fits to the data in \cref{fig:X/O-vs-Fe/C} as a function of solar activity.
It shows that these exponents may change in a cyclic fashion with solar activity.
However, data from further solar cycles are necessary to increase confidence in such an inference.

\cref{fig:X/O-vs-Fe/C-exponents-MpQ} plots \added{and compares} these exponents as a function of \MpQ\ along with the analogous trend for \replaced{LSEP}{GSEP at \MeVnuc[0.38]} events from \citet{Desai2006b}.
\deleted{It compares the \MpQ\ fractionation trend in the quiet time suprathermals studied in this paper with the analogous trend in LSEP events.}
Although the y-intercepts are different, the slopes are consistent across solar activity.
\added{As \citepossessive{Desai2006b} observations of individual GSEP events cover one solar cycle, our observations suggest that event-to-event variability in their GSEP events is consistent with the long-term variability of ST abundances in quiet times.}

\cref{fig:Fractionation-Summary} takes fits to these exponents as a function of \MpQ\ and plots the resulting slopes as a function of solar activity.
It shows that the fit uncertainty is markedly larger than the variability of our results due to \QT\ threshold selection, suggesting that the variability due to \QT\ threshold in \sect{QT-def} does not impact our results and they are robust to \QT\ selection threshold.


\subsection{Annual Abundance Correlation with SSN: Suprathermal \Su\ is low \FIP \label{sec:disc:annual-ab-SSN}}

\citet{Reames2018a,Reames2018b} argue that the difference in CIR and SEP abundances is a result of the solar source regions from which their ions \replaced{originate}{are released into the heliosphere}, in particular the first ionization potential (\FIP) effect.
The \FIP\ effect is the observation that the abundance of low \FIP\ ions is enhanced relative to their photospheric abundances and high \FIP\ ions are depleted \citep{LR-FIP}.
\citet{Reames2018b} argue that this low/high \FIP\ separation occurs at \eV[\sim 10] in SEP events and \Su\ is a high \FIP\ SEP element.
In particular, \citet{Reames2018a} argues that \Su\ behaves like higher \FIP\ elements, \replaced{like \C\  in solar wind, CIRs, and SEP events}{of which \C\ is commonly treated as a genetic example}.
As such, we would expect \AbSEP[\Su] to follow the same temporal variations as \AbSEP[\C].

This work aggregates ST measurements into annual bins, which are orders of magnitude longer in duration than events that accelerate SEPs or the processes related to injection, acceleration, and transport.
\cref{tbl:Ab-SSN-xcorr} gives the \deleted{cross} correlation coefficient \xcorr[\AbSEP,\SSN].
Given the annual duration of our ST aggregations, these \xcorr\ are necessarily related to either how the source from which the ST \replaced{pool}{population} is drawn or how the prevalence of different acceleration, injection, and transport mechanisms changes with solar activity.
As such, we can use them to test the results of \citet{Reames2018a,Reames2018b}.

\plotAnnualAbundanceXcorr
\cref{fig:annual-ab-xcorr} plots the color-coded correlation coefficient \xcorr[\AbSEP,\SSN] as a function of SEP \MpQ\ and \FIP. 
\deleted{We have confirmed prior results that ST ions are CIR remnants during solar minima and provided compelling evidence that they are consistent with SEP abundances during solar maxima.}
Many acceleration and transport processes that impact the solar wind and SEPs depend on \MpQ. 
If changes in the relative occurrence of these processes drive changes in \xcorr, then \xcorr\ should be ordered by \MpQ.
If changes in solar source regions drive changes in \xcorr, then \FIP\ should order it.
The clear change in sign \xcorr\ at $(\MpQ, \FIP) \approx (2.6, 11)$ suggests that at least one of \MpQ\ and \FIP\ is significant to long term changes in \xcorr[\AbSEP,\SSN].
The question is: which?
First, we will show that there is insufficient evidence that \xcorr\ depends on \MpQ.
Then we will discuss that changes in source regions may be related to changes in \xcorr.

Let us assume that \xcorr\ is tied to changes in the prevalence of a rigidity-dependent process.
Then \xcorr\ should be ordered by \MpQ.
Two observations indicate this is not the case.
\begin{inparaenum}[(1)]
\item Ions with $\FIP < \eV[11]$ all have $\xcorr > 0.8$, without a clear ordering.
\item High \FIP\ ions show no defined ordering.
\end{inparaenum}
While only there are only three points high \FIP\ points (\C, \N, and \Ne), we can exclude \AbSEP[\C] and its anti-correlation because \C\ is known to be overabundant in the solar wind and CIRs.
\added{In other words, \C\ is not necessarily a generic example of high \FIP\ elements.}
The two remaining points (\N\ and \Ne) are insufficient to draw any conclusion about the impact of changes in a rigidity-dependent mechanism on high \FIP\ ions.
In short, there is insufficient evidence for an \MpQ-dependent process driving \xcorr.

Let us assume the opposite, that changes in \xcorr\ are tied to source region effects at the Sun.
The \FIP\ effect yields abundance differences based on how the combination of source region temperatures and height in the solar atmosphere determine when an element ionizes.
These temperatures and heights are related to solar source region type (e.g.~coronal hole, active region, streamer belt, etc.) from which the ions emanates.
Because \FIP\ is tied to source region, this suggests a discrete change in \AbSEP\ are possible based on the source region driving each abundance.
Given the change in \xcorr's sign occurs at $\FIP \approx \eV[11]$ and there is no other ordering as a function of \MpQ, \cref{fig:annual-ab-xcorr} implies that \xcorr\ is driven by source region changes and not the prevalence of any \MpQ-dependent process.
Absent some other known physical mechanisms, that the change in \xcorr's sign occurs at \eV[11] implies that \Su\ is a low \FIP\ ST ion.
Given we have shown that STs are consistent with GSEP and ISEP populations, we must infer that \Su\ is \deleted{unexpectedly \citep{Reames2018a,Reames2018b}} low \FIP\ in SEPs--not high \FIP \added{ \citep{Reames2018a,Reames2018b}}--and that ST \Su\ must originate in the same regions as other low \FIP\ elements like \Ca, \Fe, \Mg, and \Si.


\section{Conclusion \label{sec:conclusion}} 
We have developed a method for analyzing long term trends in \MeVnuc[0.3] to \MeVnuc[1.28] ions during quiet times that includes tests for the sensitivity of our results to our quiet time selection criteria.
We refer to these as suprathermals and have shown that the statistical uncertainty in our results dominates any uncertainty due to our quiet time selection criterion.
As such, our results and \added{\QT\ threshold proposed by} \citepossessive{Dayeh2017} \deleted{(mean, variance) quiet time identification criterion }are robust.

We have confirmed that the annual number of quiet hours decreases with increasing solar activity \citep{Dayeh2017}. 
This is likely because the occurrence of \replaced{SEP accelerating events}{phenomena like flares and CMEs that accelerate SEPs} increases with \SSN.
Our results also support prior conclusions that the populations from which ST ions are drawn change with solar activity \citep{Desai2006a,Dayeh2009,Dayeh2017,Zeldovich2011,Zeldovich2018,Zeldovich2021}.
\added{In particular, normalizing quiet time \AbSEP\ to CIR abundances during solar minima and GSEP abundances during solar maxima removes the trends with \MpQ.
This suggests that the quiet time suprathermals we observe are remnants of material that was previously accelerated by the dominant energetic particle producer during that epoch, that these energetic particles decay in intensity to the suprathermal regime, and that they are not accelerated out of the solar wind nor decelerated from SEPs.
That these observations are made during quiet times further substantiates observations that suprathermals suffuse the interplanetary medium \citep{Tsurutani1985b,Desai2003,Wiedenbeck2003}.}

Coincident to these broad findings, our analysis has also \deleted{unexpectedly}revealed that suprathermal \Su\ is a low \FIP\ ion.
Given that high and low \FIP\ ions are from distinct sources on the Sun, this means that \Su\ must be treated like a low \FIP\ ion when tracing SEP events back to their solar sources and modeling SEP acceleration mechanisms.


\begin{acknowledgments}
The authors thank Heather Elliot and Frederic Allegrini feedback on this paper.
BLA acknowledges NASA grants 80NSSC20K1255 and 80NSSC21K0112.
MID acknowledges NASA grants 80NSSC20K1255, 80NSSC21K0112, and 80NSSC21K1572.
MAD acknowledges NASA grants 80NSSC19K0079 (LWS), 80NSSC20K0290 (O2R), and contract NNJ15HK11B.
GMM acknowledges 80NSSC22K0374.
\end{acknowledgments}

\software{
IPython \citep{IPython}, 
Jupyter \citep{JupyterNB}, 
Matplotlib \citep{matplotlib}, 
Numpy \citep{NumPy}, 
SciPy \citep{SciPy},
Pandas \citep{pandasA,pandasB}, 
Python \citep{pythonA,pythonB}
}

\KnownAbundanceTable
\XvsFeExponentTable
\AnnualAbundanceTable

\bibliography{Mendeley.bib}{}
\bibliographystyle{aasjournal}

\end{document}